%% file: main.tex
\documentclass[11pt,a4paper]{article}
\usepackage[utf8]{inputenc}
\usepackage[T1]{fontenc}
\usepackage[margin=1in]{geometry}
\usepackage{cite}
\usepackage{amsmath,amssymb,amsfonts}
\usepackage{subcaption}
\usepackage{booktabs}
\usepackage{float}
\usepackage{algorithmic}
\floatstyle{ruled}
\newfloat{algorithm}{tbp}{loa}
\floatname{algorithm}{Algorithm}
\usepackage{graphicx}
\usepackage{textcomp}
\usepackage{xcolor}
\usepackage{url}
\usepackage{acronym}
\usepackage[most]{tcolorbox}
\usepackage[normalem]{ulem}
\usepackage{enumitem}
\usepackage{tikz}
\usepackage{eso-pic}
\usepackage{authblk}

\definecolor{lightergray}{gray}{0.90}
\newcommand\greybox[1]{%
  \par\noindent\colorbox{lightergray}{ 
    \begin{minipage}{\dimexpr\linewidth-2\fboxsep\relax}#1\end{minipage}%
  }%
  \vskip\baselineskip%
}

\newcommand{\IEEEsubmittednotice}{%
  This work has been submitted to the IEEE for possible publication.
  Copyright may be transferred without notice, after which this version
  may no longer be accessible.%
}

\AddToShipoutPictureFG{%
  \begin{tikzpicture}[remember picture,overlay]
    \node[
      anchor=south,
      draw=black,
      line width=0.3pt,
      rounded corners=1pt,
      fill=white,
      inner xsep=4pt,
      inner ysep=2pt,
      text width=0.92\paperwidth,
      align=center,
      font=\scriptsize
    ] at ([yshift=5mm]current page.south) {\IEEEsubmittednotice};
  \end{tikzpicture}%
}

\definecolor{grpStable}{HTML}{2CA02C}
\definecolor{grpModerate}{HTML}{FF7F0E}
\definecolor{grpVolatile}{HTML}{D62728}
\definecolor{ablTotal}{HTML}{1F77B4}
\definecolor{ablEpi}{HTML}{D62728}
\definecolor{ablAle}{HTML}{2CA02C}
\newcommand{\grouplegend}{\textcolor{grpStable}{$\bullet$}~Stable\quad
  \textcolor{grpModerate}{$\bullet$}~Moderate\quad
  \textcolor{grpVolatile}{$\bullet$}~Volatile}
\newcommand{\ablationlegend}{\textcolor{ablTotal}{$\bullet$}~Total
  ($\sigma^2_{\mathrm{epi}}{+}\sigma^2_{\mathrm{ale}}$)\quad
  \textcolor{ablEpi}{$\blacksquare$}~Epistemic only ($\sigma^2_{\mathrm{epi}}$)\quad
  \textcolor{ablAle}{$\blacktriangle$}~Aleatoric only ($\sigma^2_{\mathrm{ale}}$)}
\definecolor{mixRealistic}{HTML}{1F77B4}
\definecolor{mixLowEnd}{HTML}{FF7F0E}
\definecolor{mixPredomLow}{HTML}{D62728}

\definecolor{chMild}{HTML}{2CA02C}
\definecolor{chModerate}{HTML}{FF7F0E}
\definecolor{chSevere}{HTML}{D62728}
\definecolor{chProlonged}{HTML}{9467BD}

\acrodef{LSTM}[LSTM]{Long Short-Term Memory}
\acrodef{MEC}[MEC]{Multi-access Edge Computing}
\acrodef{ETSI}[ETSI]{European Telecommunications Standards Institute}
\acrodef{VEC}[VEC]{Vehicular Edge Computing}
\acrodef{BNN}[BNN]{Bayesian Neural Network}
\acrodef{SMART}[SMART]{Strategic Mobility and Resource Uncertainty Tracker}
\acrodef{TOPS}[TOPS]{Tera Operations Per Second}
\acrodef{VRAM}[VRAM]{Video RAM}
\acrodef{CDF}[CDF]{Cumulative Distribution Function}
\acrodef{SVI}[SVI]{Stochastic Variational Inference}
\acrodef{SAA}[SAA]{Sample Average Approximation}
\acrodef{DRL}[DRL]{Deep Reinforcement Learning}
\acrodef{IMT}[IMT]{International Mobile Telecommunications}
\acrodef{RAN}[RAN]{Radio Access Network}
\acrodef{AoI}[AoI]{Area of Interest}
\acrodef{AMS}[AMS]{Application Mobility Service}
\acrodef{MC}[MC]{Monte Carlo}
\acrodef{Mp1}[Mp1]{MEC reference point between MEC application and MEC platform}
\acrodef{ELBO}[ELBO]{Evidence Lower Bound}
\acrodef{BS}[BS]{Base Station}
\acrodef{PICP}[PICP]{Prediction Interval Coverage Probability}
\acrodef{KM}{Knowledge Manager}
\acrodef{DM}{Decision Manager}
\acrodef{MAE}[MAE]{Mean Absolute Error}
\acrodef{CVaR}[CVaR]{Conditional Value-at-Risk}
\acrodef{ECE}[ECE]{Expected Calibration Error}

\def\BibTeX{{\rm B\kern-.05em{\sc i\kern-.025em b}\kern-.08em
    T\kern-.1667em\lower.7ex\hbox{E}\kern-.125emX}}
\begin{document}

\setlength{\textfloatsep}{6pt plus 1pt minus 2pt}
\setlength{\floatsep}{6pt plus 1pt minus 2pt}
\setlength{\intextsep}{6pt plus 1pt minus 2pt}
\setlength{\abovecaptionskip}{2pt}
\setlength{\belowcaptionskip}{2pt}

\title{Beyond Car Sharing: Uncertainty-Aware Pooling of Vehicular Compute at the Network Edge
}

\author[ ]{Wellington Lobato}
\author[ ]{Nadjib Achir}
\author[ ]{Aline Carneiro Viana}
\affil[ ]{\textit{Inria Saclay, France}}
\affil[ ]{\texttt{\{viana-lobato-junior.wellington, nadjib.achir, aline.viana\}@inria.fr}}

\maketitle

\begin{abstract}
Connected vehicles increasingly embed AI accelerators, offering a substantial yet volatile source of supplemental compute near the network edge. Unlike provisioned MEC hosts, vehicular resources are highly dynamic: vehicles may leave the cell, become locally occupied, or offer heterogeneous compute capacities. Therefore, exploiting vehicular resources requires making admission decisions without knowing the compute capacity that will be available during task execution.  
We present SMART, an uncertainty-aware admission mechanism that enables an \acs{ETSI} \ac{MEC} orchestrator \textit{to opportunistically exploit vehicular compute under predictive uncertainty.} 
SMART predicts future vehicular capacity using a \ac{BNN}~-- whose uncertainty estimates are the best calibrated among the evaluated forecasters at the nominal 95\% level, and incorporates its calibrated predictive uncertainty into a chance-constrained admission program reformulated through \ac{SAA} and \ac{CVaR} approximations.
Under a compute-only admission model, SMART admits 95.6\% of tasks while maintaining a median capacity-violation rate of about 0.81\%. It achieves a favorable admission--violation tradeoff compared with seven reactive, mean-only, and uncertainty-aware baselines, by approaching the performance of a compute-capacity oracle under the modeled assumptions. 
Finally, the sensitivity analyses show that variability in base-station compute availability is a key determinant of admission performance, highlighting the need for a calibrated admission and resource allocation mechanism tailored to opportunistic base-station compute pooling.
\end{abstract}


\input{01_introduction}

\input{02_background}
\input{03_system}

\input{04_evaluation}
\input{06_conclusions}

\bibliographystyle{IEEEtran}
\bibliography{reference}

\end{document}

%% file: 01_introduction.tex
\section{Introduction}
\label{sec:introduction}
\ac{ETSI} \ac{MEC} offers a standardized execution environment for deploying latency-sensitive services at the network edge~\cite{etsi2025mec}.
A \ac{MEC} host provides a fixed compute capacity, whereas computational demand fluctuates with vehicle density throughout the day, leading to saturation during peak traffic and underutilization during off-peak periods~\cite{liu2021vec}. 
At the same time, modern vehicles increasingly embed AI accelerators rated at 20--275~\ac{TOPS}, whose capacity often goes underutilized or idle during routine driving, parking, or charging~\cite{lu2023vehiclecomputing}.
Collectively, these underutilized resources can rival a co-located \ac{MEC} server in aggregate compute capacity~\cite{fan2022jointvec}.

Recent work leverages vehicles as edge compute hosts and migrates workloads when they leave the coverage area, positioning vehicle compute as a practical extension of \ac{MEC} capacity~\cite{feraudo2023,lu2023vehiclecomputing}.
However, these approaches inherit two assumptions from conventional \ac{ETSI} \ac{MEC} orchestration that do not hold in vehicular environments.
First, the orchestrator admits tasks only after observing the available resource pool, preventing admission from exploiting resources that are not yet observable and forcing operators either to overprovision the static \ac{MEC} tier or to rely on costly workload migration~\cite{feraudo2023}. Second, forecasts of the future resource pool are typically reduced to point estimates, causing admission decisions to treat these predictions as exact~\cite{nguyen2023,zhan2020} without distinguishing epistemic uncertainty, arising from incomplete knowledge, from aleatoric uncertainty, arising from inherent system variability~\cite{kendall2017uncertainties}. 
Addressing these limitations requires incorporating predictive uncertainty, i.e., both its magnitude and type, into admission decisions under partial system observability. 
This raises the following question: \textit{How should volatile vehicular capacity be committed under predictive uncertainty?}

This paper introduces \textit{\ac{SMART}, an \ac{ETSI}-compliant mechanism for uncertainty-aware admission of opportunistic vehicular compute.} 
Rather than proposing a new orchestration architecture, \ac{SMART} augments an existing \ac{ETSI} \ac{MEC} orchestrator with a probabilistic admission layer that translates predictive uncertainty into risk-aware decisions~\cite{etsi2025mec}.
By doing so, it enables admission decisions that explicitly account for uncertainty in future resource availability. \ac{SMART} comprises two functions. 
The \textit{\ac{KM}} learns the predictive distribution of future vehicular capacity using a \ac{BNN}, separating epistemic from aleatoric uncertainty. 
The \textit{\ac{DM}} propagates this distribution into a chance-constrained admission program, where a configurable risk level bounds the probability of capacity violations.

Unlike prior uncertainty-aware approaches, which typically target stable cloud environments or model uncertainty only at a coarse level~\cite{zhou2023aquatope}, \ac{SMART} directly incorporates calibrated predictive uncertainty into admission decisions for highly dynamic vehicular resource pools. 
\textit{To preserve deployability}, it reuses existing \ac{ETSI} \ac{MEC} reference points and the \ac{AMS} for departure-driven migration, introducing no additional interfaces~\cite{etsi2025mec,etsi2022ams}. 
Workload arrivals are treated as exogenous, allowing the mechanism to isolate uncertainty in resource availability. Overall, we make three classes of contributions: 

\noindent\textbf{Methodological:} 
After a detailed literature review (cf. \textbf{\S \ref{sec:background}}), we present \ac{SMART}, a mechanism for uncertainty-aware pooling of vehicle compute at the network edge (cf. \textbf{\S \ref{sec:our_work}}), along with the associated \ac{ETSI} \ac{MEC} context. The section \textbf{\S \ref{sec:our_work}} further details the two manager functions of \ac{SMART}, the \ac{BNN}-based forecasting pipeline and the chance-constrained admission program. Together, these components enable \ac{SMART} to explicitly propagate predictive uncertainty into admission decisions (cf. Fig.~\ref{fig:overview}). 
To the best of our knowledge, \ac{SMART} is the first mechanism that jointly integrates \ac{ETSI}-compliant vehicular resource pooling, probabilistic resource forecasting with separated epistemic and aleatoric uncertainty, and risk-aware admission control under explicit uncertainty constraints.

\noindent\textbf{Empirical:} For empirical evaluation, we use a pseudonymized real-world taxi mobility dataset (T-Drive~\cite{yuan2010tdrive}), containing $10\,223$ taxis and $2\,161$ \acp{BS} covering $1\,254$~km$^2$ of Beijing, China.
We implement a \ac{SMART}-embedded ETSI-compliant \ac{MEC} mechanism and evaluate it against a compute-capacity oracle and seven admission baselines, including reactive, mean-based, and uncertainty-aware approaches. 

\noindent\textbf{Code release:} The above contributions compose \ac{SMART} code~-- \textit{an uncertainty-aware admission mechanism for opportunistic vehicular compute}~-- a Python implementation to be released upon acceptance of the paper. We conclude this paper in section \textbf{\S \ref{sec:conclusion}}.


%% file: 02_background.tex
\section{The MEC-vehicle Landscape}
\label{sec:background}
Vehicular edge capacity varies with mobility and thus cannot be modeled as a stable resource pool, as is often assumed in conventional \ac{MEC} orchestration.
Vehicles may leave the cell, decline to participate, throttle under thermal constraints, reclaim accelerators for local workloads, or offer heterogeneous hardware budgets.
Exploiting vehicular pooling is therefore not only a matter of resource discovery but of allocation under uncertainty. 
This section positions \ac{SMART} with respect to vehicular far-edge resources, \ac{ETSI} compliance, and uncertainty-aware allocation.

\subsubsection{Urban Vehicular Compute Pooling at the Edge} 
\label{sec:bg_resources}

\begin{figure*}[t]
  \centering
   \begin{subfigure}[t]{0.45\linewidth}
      \centering
      \includegraphics[width=\linewidth]{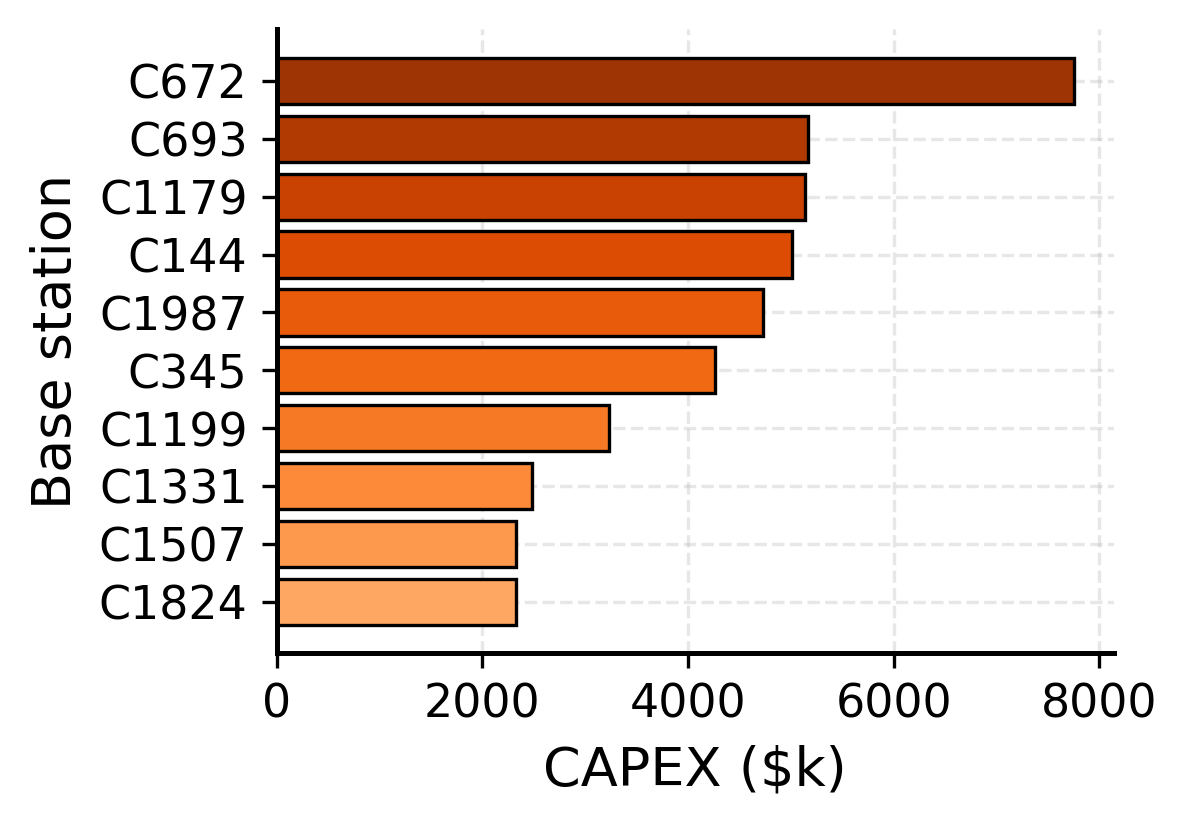}
      \caption{CAPEX of BS's peak demand.} 
      \label{fig:capex}
  \end{subfigure}
   \begin{subfigure}[t]{0.45\linewidth}
      \centering
      \includegraphics[width=0.95\linewidth]{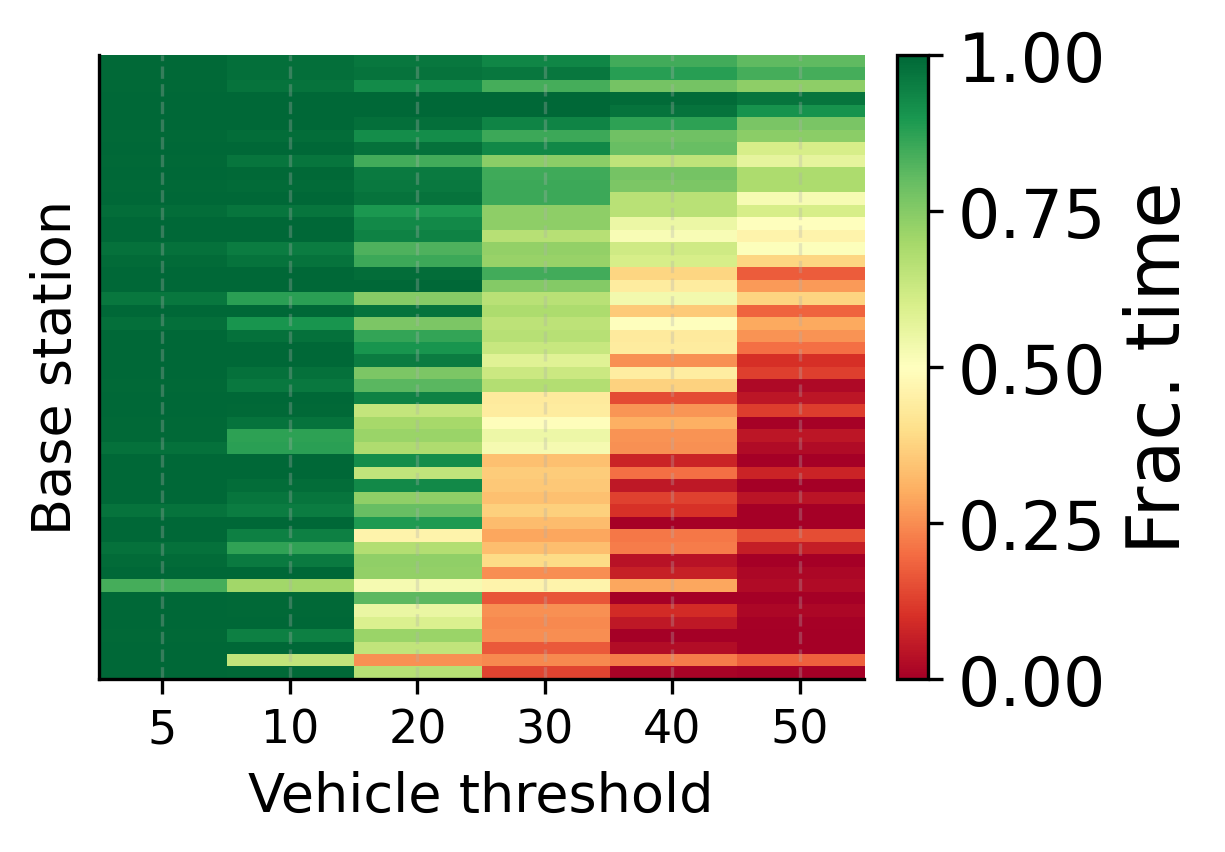}
      \caption{Time fraction per BS.}
      \label{fig:availability}
  \end{subfigure}
\begin{subfigure}[t]{0.6\linewidth}
    \centering
     \includegraphics[width=0.95\linewidth]{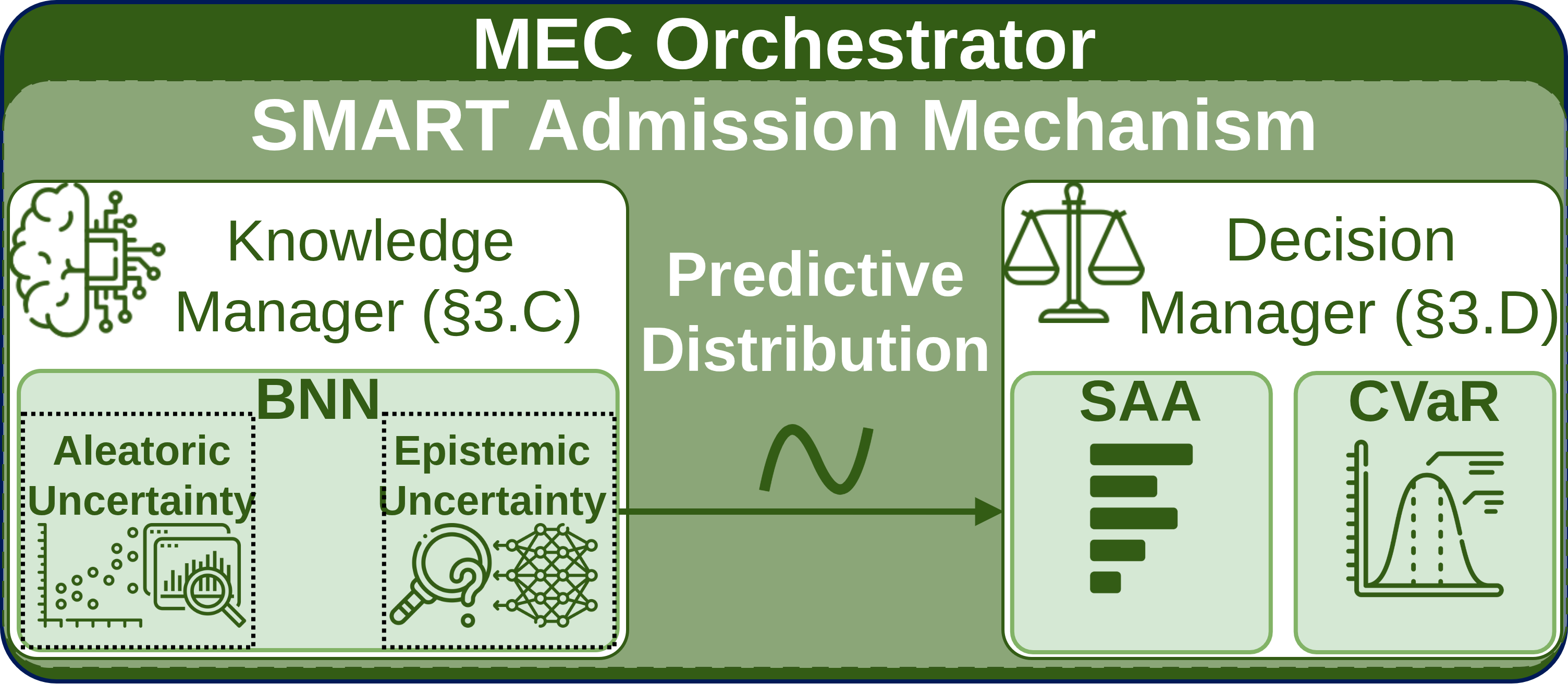}
   \caption{SMART overview.}
    \label{fig:overview}
\end{subfigure}
  \caption{(a) Capital cost of fixed accelerator deployments matching the peak vehicular presence in BSs. Provisioning fixed hardware to meet peak demand is costly and may lead to resource underutilization, as (b) high vehicle counts are sustained for only a small fraction of time. (c) SMART overview with its functional models.}
  \label{fig:availability_capex}
\end{figure*}

Manufacturers equip vehicles with sufficient computational capacity to satisfy peak processing demands (cf. Fig.~\ref{fig:capex}). 
However, autonomous navigation, sensor fusion, and other time-critical functions execute only intermittently, leaving substantial processing resources idle during normal operation~\cite{patane2025vcc} (cf. Fig.~\ref{fig:availability}). 
Fig.~\ref{fig:capex} illustrates the CAPEX required to provision fixed infrastructure per \ac{BS} to sustain peak vehicular demand, with C672 incurring the highest cost and other \acp{BS} exhibiting lower investment. 
Fig.~\ref{fig:availability} further shows that peak demand is transient and sustained only for a small fraction of the time across most \acp{BS}, confirming a persistent mismatch between provisioned capacity and typical demand. 

When aggregated across vehicles within a single \ac{BS} coverage area, idle capacity can match or exceed that of a conventional \ac{MEC} server~\cite{patane2025vcc}.
This aggregated capacity does not constitute a stable infrastructure but instead forms a transient resource pool driven by daily mobility patterns (cf. Fig.~\ref{fig:availability}). Vehicle density varies across diurnal and weekly cycles, with pronounced morning peaks and increased weekday activity~\cite{yuan2010tdrive}. 
Spatially, vehicles concentrate in commercial districts during working hours, in residential areas in the evening, and at event venues during large gatherings.
These patterns induce continuous variation in the available compute envelope within each \ac{BS}~\cite{patane2025vcc}.
Such temporal and spatial volatility complicates orchestration: decisions based on instantaneous observations risk acting on outdated system states, while single-point forecasts lack robustness under distributional shifts.

\vspace{.3cm}\greybox{\textbf{Insight:} \textit{The vehicular resource pool enables reuse of distributed computational capacity in user-dense environments, reducing the need for dedicated \ac{MEC} infrastructure~\cite{patane2025vcc}. The main trade-off is runtime orchestration under highly time-varying resource availability.}}

\subsubsection{Literature Limitations}
\label{sec:bg_limitations}
Table~\ref{tab:related_work} positions \ac{SMART} against related work.  
Existing approaches address only subsets of the design space.
None jointly integrates ETSI-compliant vehicular resource usage (cf. \textit{Vehicle usage} and \textit{ETSI compliance}), probabilistic forecasting with epistemic--aleatoric uncertainty decomposition (cf. \textit{Resource prediction} and \textit{Uncert. (e/a)}), and uncertainty-aware admission control (cf. \textit{Risk-aware}).

Among the most closely related efforts, Feraudo et al.~\cite{feraudo2023} extend \ac{ETSI} \ac{MEC} by incorporating vehicular nodes, including parked far-edge hosts, into the resource pool through participation incentives and transparent migration. Their orchestrator operates on the currently observed pool using a reactive policy, without forecasting future capacity or modeling mobility-induced uncertainty in admission decisions.

The remaining approaches address only parts of this design space.
Giannone et al.~\cite{giannone2020} assume fixed and known resource capacities; Fan et al.~\cite{fan2022jointvec} optimize vehicular offloading under known resources without \ac{ETSI} compliance; and AQUATOPE~\cite{zhou2023aquatope} introduces uncertainty awareness but targets stable cloud environments without separating epistemic and aleatoric uncertainty. 
Likewise, \ac{DRL}-based offloading frameworks generally assume that future resource states are known at decision time, relying on point estimates of system dynamics~\cite{nguyen2023}.
Mobility-aware schedulers behave analogously, forecasting individual trajectories or channel values with no uncertainty representation~\cite{zhan2020}.
Both classes of approach treat predictions as exact, failing to separate epistemic uncertainty (from incomplete knowledge) from aleatoric uncertainty (from inherent system variability)~\cite{kendall2017uncertainties}.

Taken together, these limitations reveal a fundamental missing capability: uncertainty-aware admission control over volatile vehicular resources while preserving \ac{ETSI} compliance. 
From a deployment perspective, \ac{ETSI} compliance is essential, as operators are unlikely to adopt mechanisms that require non-standard \ac{MEC} reference points.

\noindent\textbf{Positioning:} \ac{SMART} addresses the literature gaps by providing the uncertainty-aware admission logic missing from existing pool-based designs while remaining fully compliance with the \ac{ETSI} \ac{MEC} architecture. 

The \ac{SMART} admission mechanism also differs from prior uncertainty-aware decision-making approaches. Chance-constrained and robust formulations (e.g., Gaussian-process MPC, distributionally robust optimization, and conformal prediction) typically target stable cloud workloads, worst-case ambiguity sets, or distribution-free guarantees rather than highly dynamic vehicular resource pools.
In contrast, \ac{SMART} integrates Bayesian capacity forecasting, epistemic--aleatoric uncertainty decomposition, and chance-constrained admission within an \ac{ETSI}-compatible \ac{MEC} orchestration path.

\begin{table}[t]
\centering
\caption{Comparative overview of \ac{SMART} and related work.
\checkmark\,=\,addressed; \textemdash\,=\,not addressed;
$\sim$\,=\,partial.}
\label{tab:related_work}
\footnotesize
\setlength{\tabcolsep}{4pt}
\begin{tabular}{cccccc}
\hline
Work & Vehicle & ETSI & Resource & Uncert. & Risk- \\
     & usage & compliance & prediction & (e/a) & aware \\
\hline
Giannone~\cite{giannone2020}                  & \textemdash & \checkmark  & \textemdash & \textemdash & \textemdash \\
Feraudo~\cite{feraudo2023}     & \checkmark  & \checkmark  & \textemdash & \textemdash & \textemdash \\
Fan~\cite{fan2022jointvec}                    & \checkmark  & \textemdash & \textemdash & \textemdash & \checkmark  \\
Nguyen~\cite{nguyen2023}                      & \textemdash & \textemdash & \textemdash & \textemdash & \checkmark  \\
Zhan~\cite{zhan2020}                          & \textemdash & \textemdash & \checkmark  & \textemdash & \checkmark  \\
AQUATOPE~\cite{zhou2023aquatope}              & \textemdash & \textemdash & \checkmark  & $\sim$      & \checkmark  \\
\ac{SMART} (ours)                             & \checkmark  & \checkmark  & \checkmark  & \checkmark  & \checkmark  \\
\hline
\end{tabular}
\end{table}

\vspace{.3cm}\greybox{\textbf{Insights:} \textit{Deployable vehicular \ac{MEC} systems must preserve strict \ac{ETSI} conformance while explicitly handling uncertainty in highly dynamic vehicular resource environments. This, in turn, requires propagating calibrated predictive distributions into admission decisions rather than relying on deterministic resource forecasts.}}\vspace{-0.3cm}

%% file: 03_system.tex
\section{The SMART Admission Mechanism} 
\label{sec:our_work}


\ac{SMART} extends the \ac{ETSI} \ac{MEC} orchestrator with uncertainty-aware admission mechanism for dynamic vehicular computing resources. 
Specifically, it addresses two challenges in opportunistic vehicular computing: (i) admission decisions should account for future resource availability rather than only the current state; and (ii) predicted resource availability should be treated as uncertain rather than as deterministic point estimates because of mobility-driven fluctuations.


\begin{itemize}[leftmargin=*]
    \item \textbf{Knowledge Manager (KM):} uses a \ac{BNN} to learn a calibrated predictive distribution of per-cell schedulable capacity, separating epistemic and aleatoric uncertainty (\S\ref{sec:learning});  
    \item \textbf{Decision Manager (DM):} transforms predicted distributions into risk-aware admission decisions by solving a chance-constrained optimization problem using \textit{\ac{SAA}} or \ac{CVaR} approaches (\S\ref{sec:decision}).
\end{itemize}

\ac{SMART} does not introduce any additional \ac{ETSI} reference points~\cite{etsi2025mecmanagement}. The \ac{KM} module utilizes resource-monitoring information already available within the MEC management framework, while the \ac{DM} module converts admission decisions into standard operations for the MEC application lifecycle. Thus, \ac{SMART} enhances orchestration while preserving interoperability with existing ETSI MEC deployments.

Next, we present the formalization of the vehicle compute pooling problem, followed by the \ac{KM} and the \ac{BNN} used to estimate predictive capacity distributions. Finally, we present the \ac{DM} and the corresponding risk-aware admission policies based on \ac{SAA} and \ac{CVaR} optimization. 

\subsection{Problem Formalization} 
\label{sec:system_model}



We consider a \ac{ETSI} \ac{MEC} system composed of a \textit{set of cells} $\mathcal{K}=\{1,\dots,K\}$, where each cell corresponds to the coverage area of a \ac{BS} and is managed by a local orchestrator running \ac{SMART}. Time is discretized into one-second epochs. Let $\mathcal{V}_k(t)$ denote the \textit{set of vehicles} present in \textit{cell} $k$ at \textit{time} $t$ and $N_k(t)=|\mathcal{V}_k(t)|$ the corresponding \textit{vehicle count}. 

We assume that each \textit{vehicle} $i$ allocates a portion of its computing resources to support the execution of external tasks. The available capacity depends on the embedded hardware characteristics, which we denote by the \textit{nominal budget} $c_i$ in \ac{TOPS}. A fraction, $u_i(t)\in[0,1]$, of this capacity is consumed onboard for applications such as perception, navigation, and infotainment. The \textit{residual capacity} available for external tasks is therefore given by $c_i\,(1-u_i(t))$. In a practical deployment, vehicles periodically report their currently available compute capacity to the local \ac{ETSI} \ac{MEC} orchestrator. 

In addition to the vehicles' capacities, each cell is provided with a static \ac{ETSI} \ac{MEC} host with a fixed \textit{compute budget}, denoted $c_0$. In this case, the orchestrator aggregates the reported capacities of all vehicles currently present in the cell and combines them with the capacity of the static \ac{ETSI} \ac{MEC} host to obtain the \textit{total schedulable capacity} available in cell $k$ at time $t$, which is equal to:
\begin{equation}
  C_k(t) \;=\; c_0 + \sum_{i\in\mathcal{V}_k(t)}\,c_i\,\bigl(1-u_i(t)\bigr).
  \label{eq:capacity}
\end{equation}

This \textit{real capacity} $C_k(t)$ is time-varying due to fluctuations in vehicle workloads and changes in the vehicle population $\mathcal{V}_k(t)$ caused by mobility. As a result, \textit{the amount of computing resources available to the orchestrator cannot be considered constant and must be predicted in advance.} 

Thus, the \ac{ETSI} \ac{MEC} orchestrator continuously observes $C_k(t)$ and builds a \textit{historical capacity trace}.
This trace serves as input to the \textit{\ac{KM}}, which predicts the schedulable capacity over a prediction horizon $h$, noted $C_k(t+h)$.
The output of the KM predictor is \textit{a distribution that captures both the expected future schedulable capacity and the uncertainty} resulting from mobility and workload variation.

The \ac{KM} predictive distribution 
is then provided to the \textit{\ac{DM}}, whose role is to determine which incoming tasks should be admitted for execution within the cell. At each \textit{decision epoch} $t$, the \ac{DM}  receives a \textit{set of candidate tasks} $\mathcal{J}_W(t)$ to be executed during the \textit{future window} $[t,t+h]$, where $W=h$. Each task $j\in\mathcal{J}_W(t)$ is characterized by a \textit{computing demand} $w_j$ (in \ac{TOPS}). In this case, the admission decision can be represented by the binary variable $a_j \in {0,1}$ for $j \in \mathcal{J}_W(t)$, where $a_j=1$ indicates that \textit{task $j$ is admitted} for local execution and $a_j=0$ indicates that \textit{it is redirected to the cloud}. The total admitted load is equal to:
\begin{equation}
L_k(\mathbf{a}) = \sum_{j\in\mathcal{J}_W(t)} w_j a_j.
\label{eq:load}
\end{equation}

If the \textit{future real schedulable capacity} $C_k(t+h)$ were known when admission decisions are made, admission control would reduce to the following deterministic optimization problem:
\begin{equation}
\max_{\mathbf{a}\in\{0,1\}^{|\mathcal{J}_W(t)|}}
\sum_{j\in\mathcal{J}_W(t)} a_j
\quad
\text{s.t.}
\quad
L_k(\mathbf{a}) \le C_k(t+h),
\label{eq:p0}
\end{equation}

In practice, however, $C_k(t+h)$ is unknown because it depends on future vehicle availability. 
Therefore, since the \ac{KM} predicts $C_k(t+h)$ as a \textit{conditional probability distribution}, the \ac{DM} replaces the deterministic constraint in \eqref{eq:p0} with the following probabilistic chance constraint:

\begin{equation}
\begin{aligned}
\max_{\mathbf{a}\in\{0,1\}^{|\mathcal{J}_W(t)|}}
\quad &
\sum_{j\in\mathcal{J}_W(t)} a_j \\
\text{s.t.}\quad &
\Pr\left(
L_k(\mathbf{a}) \le C_k(t+h)
\right)
\ge 1-\epsilon .
\end{aligned}
\label{eq:p1}
\end{equation}

\noindent where $\epsilon\in[0,1]$ is the \textit{admissible violation probability}. \ac{DM} therefore allows the admitted load to exceed the future capacity only with probability at most $\epsilon$. \textit{The parameter $\epsilon$ therefore controls the trade-off between aggressive local admission and protection against capacity shortfalls.}

We consider two risk-aware admission approaches in the DM: (i) a \textit{Sample Average Approximation (SAA)} of the chance constraint, and (ii) a \textit{Conditional Value-at-Risk (CVaR)} criterion to account for the severity of potential capacity deficits.

\subsection{Knowledge Manager}
\label{sec:learning}


In order to predict the schedulable capacity of a cell and quantify the associated uncertainty, \ac{KM} relies on a \ac{BNN}, which produces probabilistic predictions rather than a single capacity estimate.
Since $C_k(t+h)$ denotes the \textit{schedulable capacity} to be predicted $h$ seconds ahead, and given past capacity observation as input $x_t$, we can derive the \textit{predictive distribution} using the \ac{BNN} as follow:
\begin{equation}
  p\!\left(C_k(t+h)\mid x_t,\mathcal{D}\right)
  \;=\;
  \int p\!\left(C_k(t+h)\mid x_t,\theta\right)\, p(\theta\mid\mathcal{D})\,\mathrm{d}\theta.
  \label{eq:bnn_predictive}
\end{equation}
where $\theta$ denotes the \textit{network parameters}, $\mathcal{D}$ is the \textit{dataset}, and $p(\theta|\mathcal{D})$ is the \textit{posterior distribution} of the network parameters.

A separate \ac{BNN} model is trained for each cell using its own capacity history. The objective is to preserve cell-specific mobility patterns and uncertainty characteristics, enabling the predictive distribution to be calibrated to local capacity dynamics rather than averaging across heterogeneous traffic regimes. 
The \textit{input} $x_t\in\mathbb{R}^{307}$ comprises the most recent capacity observations sampled at $1$~Hz (i.e., $300$ last seconds) with seven calendar features (time of day and day of week as $\sin$/$\cos$ pairs, plus a weekend indicator).

The input data are not used directly in their raw form. They are first normalized using the mean and standard deviation computed on the training set. This step facilitates \ac{BNN} training by bringing all variables to a comparable scale. The model, therefore, learns and produces its predictions in this normalized space. The predicted future capacities are then transformed back to the cell's original capacity scale, so that the admission controller can make decisions using values expressed in real capacity units.

We consider a \ac{BNN} model, $f_\theta:\mathbb{R}^{307}\rightarrow\mathbb{R}$, composed of two hidden layers of widths $64$ and $32$, followed by a scalar output. The model is implemented using Bayesian residual blocks to improve learning stability while enabling uncertainty quantification. Batch normalization is used within these blocks, and dropout with a rate of $0.1$ provides regularization to mitigate overfitting. 
Both model uncertainty and observation noise are captured by the \ac{BNN} using a probabilistic formulation in which the network parameters are assigned prior distributions, and the output is modeled through a Gaussian likelihood. The \textit{posterior distribution} of the model parameters is approximated using \ac{SVI}. At inference time, the \ac{KM} draws \textit{$S$ samples} from the approximated posterior distribution and generates \textit{$S$ corresponding predictions} of the future capacity. Finally, the predictive mean capacity is then estimated as
\begin{equation}
  \hat{\mu}(x_t)
  =
  \frac{1}{S}
  \sum_{s=1}^{S}
  f_{\theta^{(s)}}(x_t).
  \label{eq:pred_mean}
\end{equation}

The uncertainty associated with the prediction is decomposed into: (i) \textit{epistemic uncertainty} ($\hat{\sigma}^{2}_{\mathrm{epi}}$), reflecting the model's lack of knowledge, and (ii) \textit{aleatoric uncertainty} ($ \hat{\sigma}^{2}_{\mathrm{ale}}$), capturing the intrinsic variability of the capacity process. Together, they compose the total variance ($\hat{\sigma}^{2}_{\mathrm{tot}}(x_t)$): 
\begin{align}
  \hat{\sigma}^{2}_{\mathrm{epi}}(x_t)
  &=
  \frac{1}{S}
  \sum_{s=1}^{S}
  \Big(
  \mu^{(s)}(x_t)
  -
  \hat{\mu}(x_t)
  \Big)^2,
  \label{eq:var_epi}
  \\
  \hat{\sigma}^{2}_{\mathrm{ale}}
  &=
  \frac{1}{S}
  \sum_{s=1}^{S}
  \Big(
  \sigma^{(s)}
  \Big)^2,
  \label{eq:var_ale}
  \\
  \hat{\sigma}^{2}_{\mathrm{tot}}(x_t)
  &=
  \hat{\sigma}^{2}_{\mathrm{epi}}(x_t)
  +
  \hat{\sigma}^{2}_{\mathrm{ale}} .
  \label{eq:var_tot}
\end{align}

\noindent where $\sigma^{(s)}$ denotes the \textit{observation-noise standard deviation} associated with the $s$-th posterior sample.

The \textit{epistemic uncertainty}~\eqref{eq:var_epi} is reducible with additional data, whereas the \textit{aleatoric uncertainty}~\eqref{eq:var_ale} represents an irreducible homoscedastic floor. Their relative contributions are summarized by the \textit{epistemic-to-total variance ratio}:
\begin{equation}
  \rho(x_t)
  \;=\;
  \frac{\hat{\sigma}^{2}_{\mathrm{epi}}(x_t)}{\hat{\sigma}^{2}_{\mathrm{tot}}(x_t)}\in[0,1],
  \label{eq:rho}
\end{equation}

Finally, once de-standardized, the $S$ predictive samples forming the  \textit{forecast schedulable capacity distribution}, $\hat{C}_{k}^{(s)}(t+h)|^1_S$, 
and the respective uncertainties are sent to the \ac{DM} for the risk-aware admission (cf. \S\ref{sec:decision}). 

\subsection{Decision Manager}
\label{sec:decision}


The objective of the \ac{DM} is to determine the set of tasks that can be admitted while controlling the risk of future capacity shortages. Starting from the chance-constrained formulation introduced in \eqref{eq:p1}, we investigate two scenario-based approaches. The first approach uses \ac{SAA} to control the frequency of capacity violations, while the second incorporates \ac{CVaR} to account for the severity of such violations.

\subsubsection{\textbf{SAA-Based Admission}}

Solving the chance-constrained problem \eqref{eq:p1} requires access to the forecast schedulable capacity distribution $\hat{C}_k(t+h)$. 
Since the \ac{KM} provides only a finite set of predictive scenarios rather than a closed-form probability density, the \ac{DM} approximates the chance-constraint using the \ac{SAA} framework. The probability constraint in \eqref{eq:p1} is thus replaced by the empirical counterpart as follow: 
\begin{equation}
\frac{1}{S}
\sum_{s=1}^{S}
\mathbf{1}
\{
L_k(\mathbf{a})
\le
\hat{C}_{k}^{(s)}(t+h)
\}
\ge
1-\varepsilon,
\label{eq:saa_chance}
\end{equation}

\noindent where, $\varepsilon\in(0,1]$ denotes the \textit{admissible violation probability}. 

Since the admitted load $L_k(\mathbf{a})=\sum_{j} w_j a_j$ does not depend on the scenario index $s$, 
the empirical violation probability remains below $\varepsilon$ if and only if the admitted load does not exceed the empirical lower-tail $\varepsilon$-quantile of the scenarios.

Let consider $\hat{C}_{k}^{(1)}\!\le\!\dots\!\le\!\hat{C}_{k}^{(S)}$ denote the ordered capacity scenarios. In this case, the corresponding \textit{\ac{SAA} capacity  budget} is defined by
\begin{equation}
  \widehat{\mathcal{C}}_{k}^{\,\mathrm{SAA}}(\varepsilon)
  \;=\; \hat{C}_{k}^{(\lceil \varepsilon S\rceil)}(t+h),
  \label{eq:saa_cap}
\end{equation}

This value corresponds to the empirical lower-tail quantile, ensuring that at most an $\varepsilon$-fraction of scenarios violate the capacity constraint. Therefore, the \textit{chance-constrained admission problem} reduces to
\begin{equation}
\begin{aligned}
\text{(P1-SAA-KP)}\qquad
\max_{\mathbf{a}\in\{0,1\}^{|\mathcal{J}_W(t)|}}
&\quad \sum_{j} a_j \\
\text{s.t.}
&\quad L_k(\mathbf{a})
\le
\widehat{\mathcal{C}}_{k}^{\,\mathrm{SAA}}(\varepsilon).
\end{aligned}
\label{eq:p1_saa-kp}
\end{equation}

\vspace{0.2cm}
\greybox{\noindent\textbf{Insight:} \textit{A smaller value of $\varepsilon$ selects a lower capacity budget, resulting in a more conservative admission policy.}}

\subsubsection{\textbf{CVaR-Based Admission}} 
While the \ac{SAA} formulation controls the frequency of capacity shortages, it does not account for their severity. In practice, the operational cost of an over-admitted event typically increases with the magnitude of the resulting capacity deficit. To address this limitation, a \ac{CVaR}-based admission approach is also considered.

If we define the over-admitted loss as $\Lambda(\mathbf{a},C)=L_k(\mathbf{a})-C$, the corresponding admission problem is formulated as
\begin{equation}
\begin{aligned}
  \text{(P1-CVaR)}\quad
  \max_{\mathbf{a}\in\{0,1\}^{|\mathcal{J}_W(t)|}} \;&
  \sum_{j} a_j \\
  \text{s.t.} \;&
\mathrm{CVaR}_{\alpha}\!\big(\Lambda(\mathbf{a},C_k(t+h))\big) \le 0,
\end{aligned}
\label{eq:p1_cvar}
\end{equation}

\noindent where $\mathrm{CVaR}_{\alpha}$ is the \textit{expected loss }over the worst $\alpha$-fraction of outcomes.

Because the admitted load $L_k(\mathbf{a})$ is deterministic with respect to the capacity scenarios, 
the largest losses occur in the lowest-capacity realizations. Thus, the \ac{CVaR} of the loss can be computed directly from the worst $\alpha$-tail of $\Lambda$, which is the lowest $\alpha$-tail of the capacity scenarios. Let $S_{\alpha}^{\mathrm{low}}(k)$ denote the set containing the $\lceil \alpha S\rceil$ smallest capacity samples for cell $k$. The \ac{CVaR} constraint is equivalent to
\begin{equation}
L_k(\mathbf{a})
\le
\widehat{\mathcal{C}}_{k}^{,\mathrm{CVaR}}(\alpha),
\label{eq:cvar_constraint}
\end{equation}
\noindent where the corresponding \textit{\ac{CVaR} capacity budget} is 
\begin{equation}
  \widehat{\mathcal{C}}_{k}^{\,\mathrm{CVaR}}(\alpha)
  \;=\;
  \frac{1}{\lceil \alpha S\rceil}
  \sum_{s \in S_{\alpha}^{\,\mathrm{low}}(k)} \hat{C}_{k}^{(s)}(t+h),
  \label{eq:cvar_cap}
\end{equation}

Hence, the \textit{chance-constrained admission problem} becomes
\begin{equation}
\begin{aligned}
\text{(P1-CVaR-KP)}\quad
\max_{\mathbf{a}\in\{0,1\}^{|\mathcal{J}_W(t)|}}
&\sum_{j\in\mathcal{J}_W(t)} a_j \\
\text{s.t.}\quad
&L_k(\mathbf{a})
\le
\widehat{\mathcal{C}}_{k}^{,\mathrm{CVaR}}(\alpha).
\end{aligned}
\label{eq:p1_cvar_kp}
\end{equation}

\greybox{\noindent\textbf{Insight:} \textit{A smaller value of $\alpha$ places greater emphasis on the lowest capacities realizations, resulting in a conservative admission policy.}}

\subsubsection{\textbf{Problem resolution}}
The two formulations address different aspects of predictive uncertainty. While the \ac{SAA} budget in \eqref{eq:saa_cap} depends on a single lower-tail order statistic, the \ac{CVaR} budget in \eqref{eq:cvar_cap} averages the entire lower-capacity tail. The two budgets diverge when the predictive distribution is highly dispersed or asymmetric. This is particularly \textit{the case when epistemic uncertainty contributes significantly to the total uncertainty measured by \eqref{eq:rho}}.

Once the risk-adjusted capacity budget is determined, both \eqref{eq:p1_saa-kp} and \eqref{eq:p1_cvar_kp} admission formulations can be expressed as \textit{0/1 Knapsack Problem} over the set of candidate tasks in $|\mathcal{J}_W(t)|$. 
The \textit{capacity budget} is computed from the\textit{ $S$ predictive samples}, and the resulting knapsack problem can be solved using any standard knapsack solver; in our implementation, we use 
a \textit{branch-and-bound mixed-integer solver (CBC)}.

This work does not propose an online mechanism that adapts the \textit{risk-aware admission parameters}, $\varepsilon$ and $\alpha$, during operation. Instead, their operating regimes are characterized empirically. For each \textit{cell} $k$, $\varepsilon$ and $\alpha$ are swept over a predefined grid to solve \eqref{eq:p1_saa-kp} and \eqref{eq:p1_cvar} on the test windows. The resulting admission rate and capacity overshoot are then measured to evaluate the impact of the risk parameters. 


\vspace{0.2cm}
\greybox{\noindent\textbf{Insight:} 
\textit{The risk-aware admission parameters $\varepsilon$ and $\alpha$ govern the trade-off between task admission and the risk of capacity over-commitment. They therefore require empirical calibration (cf. \S\ref{sec:evaluation}) to identify operating points that balance resource utilization and over-commitment risk.}}

  \begin{table}[t]
  \centering
  \caption{Notation used throughout the paper.} 
  \label{tab:notation}
  \footnotesize
  \setlength{\tabcolsep}{4pt}
  \begin{tabular}{@{}l p{0.74\columnwidth}@{}}
  \toprule
  Symbol & Meaning \\
  \midrule
  \multicolumn{2}{@{}l}{\textit{System model: Base stations, vehicles, tasks, time}}\\
  $\mathcal{D}$~; $\mathcal{K}~; k$                & T-Drive dataset; set of BS; its index  \\
   $x_t$                        & input $\mathcal{D}$: $300$ capacity lags and $7$ calendar features \\
  $W~; h$                      & decision window; prediction horizon: $W{=}h{=}1$s \\ 
  $N_k(t)$                     & vehicle count: $N_k(t)=|\mathcal{V}_k(t)|$ \\
  $\mathcal{J}_W(t) ; j$       & candidate tasks for window $[t,t{+}h]$; task index \\
  $w_j$                        & compute demand of task $j$ (\ac{TOPS}) \\
  $a_j~; \mathbf{a}$           & admission variable of task $j$; admission vector \\
  $L_k(\mathbf{a})$            & admitted load: $L_k(\mathbf{a})=\sum_j w_j a_j$ \\
    \midrule
  \multicolumn{2}{@{}l}{\textit{Schedulable capacity~-- Eq.~\eqref{eq:capacity}}}\\
  $c_i~; u_i(t)$               & nominal budget; onboard utilization of vehicle $i$ \\
  $c_0$                        & static operator-owned host budget (\ac{TOPS}) \\
  $C_k(t)$                     & real schedulable capacity of BS$_k$ at time $t$ (\ac{TOPS})  \\
  $C_k(t{+}h)$                 & future real schedulable capacity of BS$_k$ at time $t{+}h$ \\ 
  \midrule
  \multicolumn{2}{@{}l}{\textit{Knowledge Manager (\ac{BNN})}}\\
  $\mu^{(s)}(x_t)~; \hat\mu(x_t)$ & per-sample ($s$) mean; predictive mean (\ref{eq:pred_mean}) \\
  $\hat{C}_k^{(s)}(t{+}h)$     & $s$-th future estimated or predicted schedulable capacity \\ \\
  \multicolumn{2}{@{}l}{\textit{Uncertainty decomposition}}\\
  $\hat\sigma^2_{\mathrm{epi}}~; \hat\sigma^2_{\mathrm{ale}}$ & epistemic (\ref{eq:var_epi}) and aleatoric (\ref{eq:var_ale}) uncertainties \\
  $\hat\sigma^2_{\mathrm{tot}}$ & total variance: $\hat\sigma^2_{\mathrm{tot}}=\hat\sigma^2_{\mathrm{epi}}{+}\hat\sigma^2_{\mathrm{ale}}$ \\
  $\rho(x_t)$                  & epistemic-to-total variance ratio (\ref{eq:rho}):
  $\rho(x_t)=\hat\sigma^2_{\mathrm{epi}}/\hat\sigma^2_{\mathrm{tot}}$ \\
  \midrule
  \multicolumn{2}{@{}l}{\textit{Decision Manager (risk-aware admission)}}\\
  $\varepsilon$                & \ac{SAA} admissible violation probability \\
  $\widehat{\mathcal{C}}_k^{\mathrm{SAA}}(\varepsilon)$ & \ac{SAA} capacity budget of the lower-tail $\varepsilon$-quantile (\ref{eq:saa_cap}) \\
  $\alpha$                     & \ac{CVaR} tail     probability level \\
  $\widehat{\mathcal{C}}_k^{\mathrm{CVaR}}(\alpha)$ & \ac{CVaR} capacity budget of the worst $\alpha$-tail (\ref{eq:cvar_cap}) \\
  $\varepsilon^\star,\,\alpha^\star$ & per-BS operating risk levels (\S\ref{sec:eval_dm}) \\
  \bottomrule
  \end{tabular}
  \end{table}

%% file: 04_evaluation.tex
\section{Evaluation}
\label{sec:evaluation}
We evaluate \ac{SMART} on vehicular mobility traces. \S\ref{sec:eval_setup} describes the dataset, the pipeline that turns raw GPS trajectories into the capacity trace, and the experimental configuration. \S\ref{sec:eval_km} then assesses the \ac{KM}'s forecasting accuracy and calibration. Finally, \S\ref{sec:eval_dm}--\S\ref{sec:eval_baselines} evaluate the \ac{DM}'s admission against a compute-capacity oracle and seven baselines.



\subsection{Evaluation Setup}
\label{sec:eval_setup}
\subsubsection{Dataset and Capacity Generation}

We use the T-Drive dataset~\cite{yuan2010tdrive}, which contains one week of GPS trajectories collected from a Beijing taxi fleet, as a realistic proxy for connected vehicles contributing as computing resources at the edge. The dataset comprises $10\,223$ vehicles and $2\,161$ \acp{BS} covering approximately $1\,254$~km$^2$ in Beijing. Because GPS observations are sparse and irregular, with position reports typically available every few minutes, the dataset provides only intermittent observations of vehicle locations.

We transform the raw trajectories into per-base-station vehicle counts. We start by restricting the GPS records only to the Beijing bounding box (latitude $[39.76,40.09]$, longitude $[116.17,116.57]$). Thereafter, we map each GPS sample to its nearest BS using a Voronoi partition of the BS locations. Finally, since the original traces are sparsely sampled, we apply a zero-order hold (ZOH) interpolation to reconstruct a one-second vehicle association process. Under this assumption, a vehicle remains attached to its last observed BS until it receives a new GPS update.

At each second, the reconstructed associations define the set of vehicles present in cell $k$, denoted by $\mathcal{V}_k(t)$, and the corresponding vehicle count $N_k(t)=|\mathcal{V}_k(t)|$. These quantities are then used to compute the schedulable capacity $C_k(t)$ introduced in Section~\ref{sec:system_model}. In addition, each vehicle contributes to a computing capacity of $275$~\ac{TOPS}, comparable to a modern automotive accelerator. To isolate the impact of vehicular resource dynamics, we assume the MEC host contributes no compute resources ($c_0 = 0$), such that the total available capacity is determined solely by the vehicles.

Finally, in order to evaluate \ac{SMART} under different mobility conditions, we select $45$ \acp{BS} that exhibit the greatest temporal variability in vehicle availability. Variability is quantified using the standard deviation of the vehicle count process $N_k(t)$. The selected \acp{BS} are sorted according to this metric and partitioned into three groups of $15$ \acp{BS}: \textit{Stable}, \textit{Moderate}, and \textit{Volatile}. 
The corresponding thresholds are defined by the $33$rd and $67$th percentiles of the vehicle-count standard deviation ($4.01$ and $10.66$ vehicles, respectively) and remain fixed throughout the evaluation: $2.68$ to $3.81$ vehicles in \textit{Stable} (median $3.40$), from $4.11$ to $4.91$ in \textit{Moderate} (median $4.57$), and from $22.16$ to $156.48$ in \textit{Volatile} (median $44.63$).

\subsubsection{Experimental Configuration}

We train a separate \ac{KM} for each BS using its own capacity history. The capacity traces are split chronologically: the first four days for training, the fifth for validation, and the sixth for testing. The validation day is used for temperature calibration and early stopping, while all reported results are obtained from the held-out test day. At each decision epoch, the \ac{KM} draws $S=100$ posterior samples of $C_k(t+h)$; \S\ref{sec:eval_km} shows that admission performance is insensitive to this sample count. Both the decision window and the prediction horizon are set to 1~s ($W=h=1$~s).

The \ac{DM} is evaluated under workloads that keep each BS close to saturation. Tasks arrive at a constant rate of $100$~tasks/s throughout the test day. The compute demands are scaled so that the average offered load matches the average schedulable capacity of each BS. This normalizes the offered-load-to-capacity ratio across BSs while preserving differences in capacity dynamics. 
Each task represents a generic \ac{VEC} inference job characterized by its input size, compute demand $w_j$, and deadline.

At each decision epoch, the \ac{DM} solves one admission problem. As shown in \S\ref{sec:decision}, both the \ac{SAA} \eqref{eq:p1_saa-kp} and \ac{CVaR} \eqref{eq:p1_cvar_kp} formulations reduce to a 0/1 knapsack problem, and are solved using the CBC branch-and-bound solver. We also considered an oracle policy that uses the true future capacity $C_k(t+h)$, thereby providing the maximum achievable admission rate under perfect knowledge. The \ac{SAA} policy uses the $\varepsilon$-quantile of the predictive scenarios, whereas the \ac{CVaR} policy uses the average of the worst $\alpha$ fraction of capacity realizations. 

In this sense, the parameters $\varepsilon$ and $\alpha$ are independently swept over the set $\{0.1,0.2,\ldots,0.9,0.99\}$ to characterize the impact of risk sensitivity. Across the $45$ BSs, this produces
$45 \times (10 + 10 + 1) = 945$ experimental runs. For each run, we measure the admission rate, admission loss with respect to the oracle, capacity overshoot, cloud offloading, and realized capacity utilization. The parameter sweep allows us to identify operating points that balance admission performance and overcommitment risk and to relate them to the uncertainty characteristics of each BS (\S\ref{sec:eval_dm}).


\subsection{KM: Predictive Performance}
\label{sec:eval_km}

\begin{figure*}[t]
    \centering
    \begin{subfigure}[t]{0.24\textwidth}
        \centering
        \includegraphics[width=\linewidth]{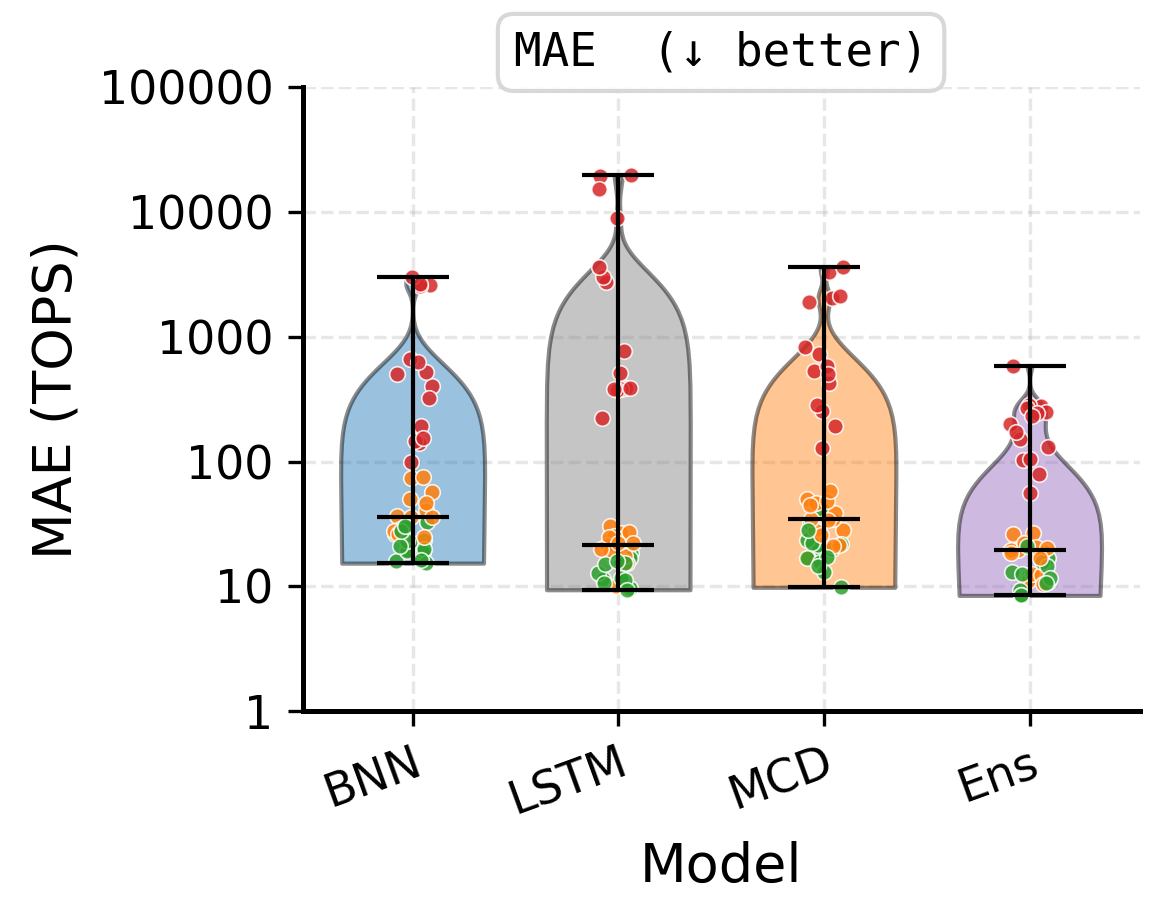}
        \caption{Mean absolute error.}
        \label{fig:km_mae}
    \end{subfigure}
    \hfill
    \begin{subfigure}[t]{0.24\textwidth}
        \centering
        \includegraphics[width=\linewidth]{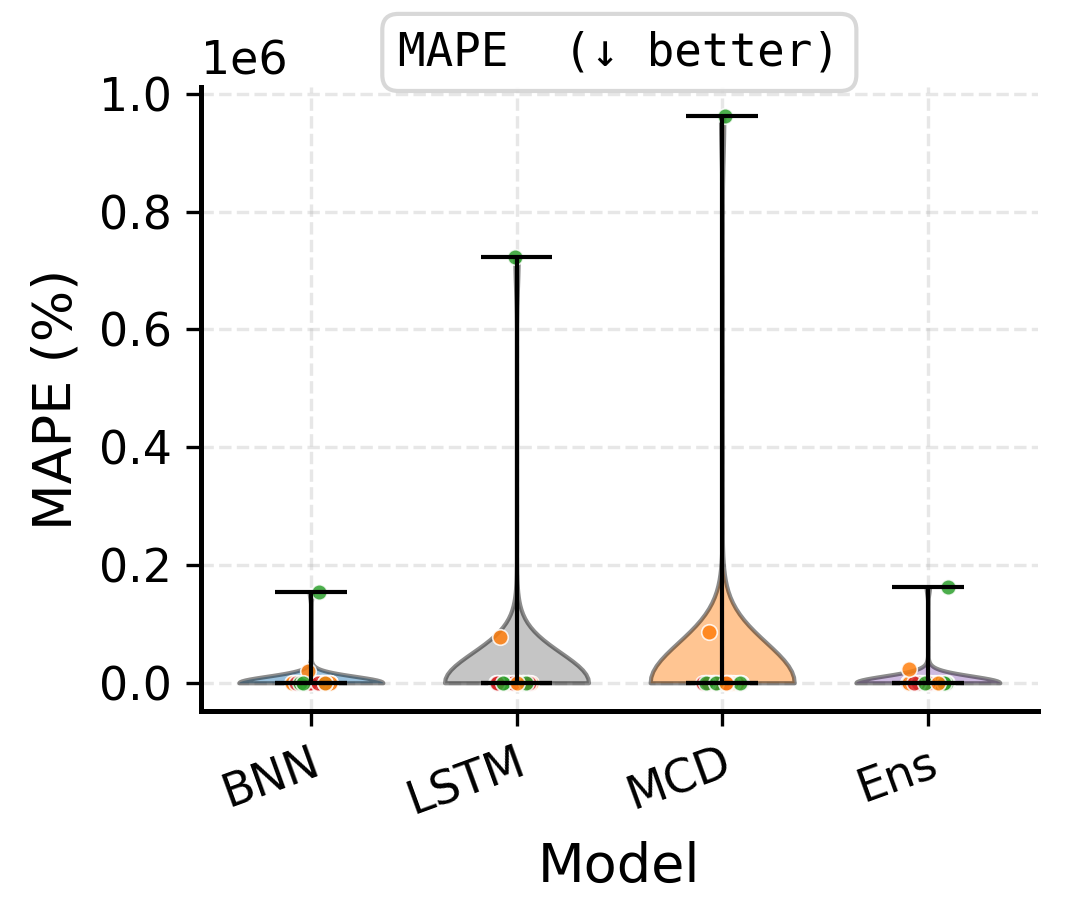}
        \caption{Mean absolute percentage error.}
        \label{fig:km_mape}
    \end{subfigure}
    \hfill
    \begin{subfigure}[t]{0.24\textwidth}
        \centering
        \includegraphics[width=\linewidth]{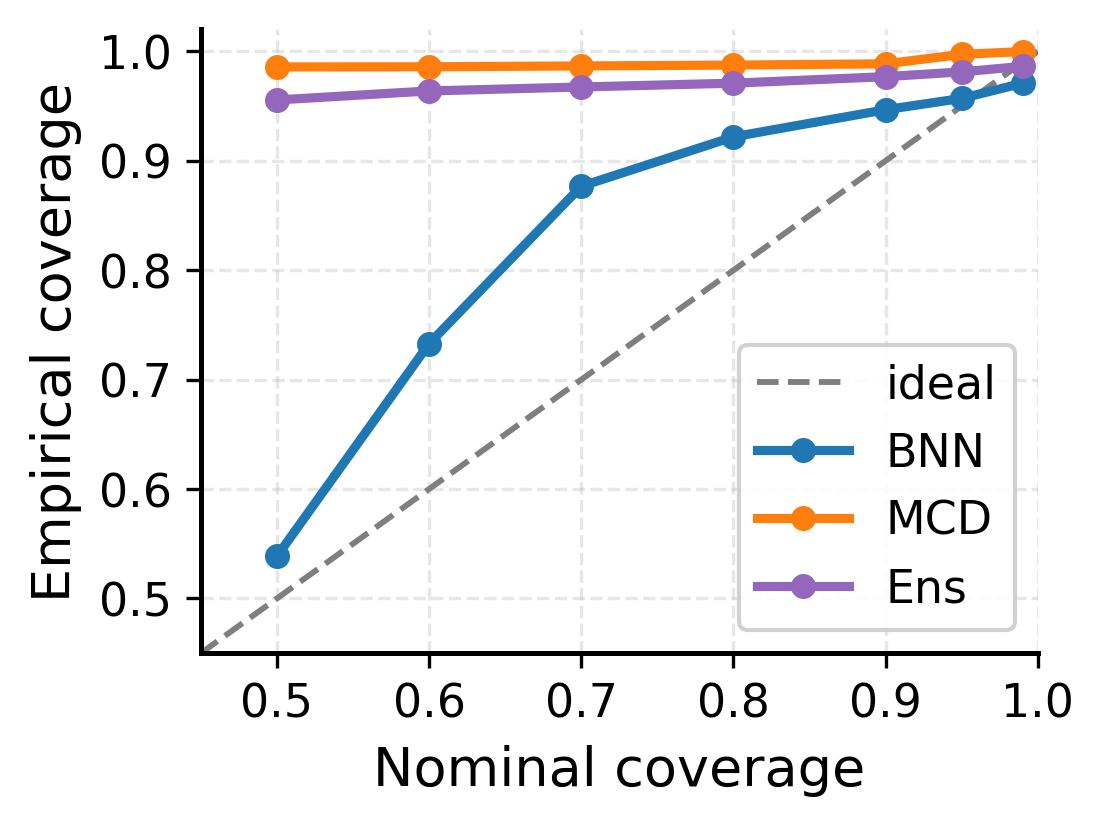}
        \caption{Reliability vs.\ model.}
        \label{fig:km_reliability}
    \end{subfigure}
    \hfill
    \begin{subfigure}[t]{0.24\textwidth}
        \centering
        \includegraphics[width=\linewidth]{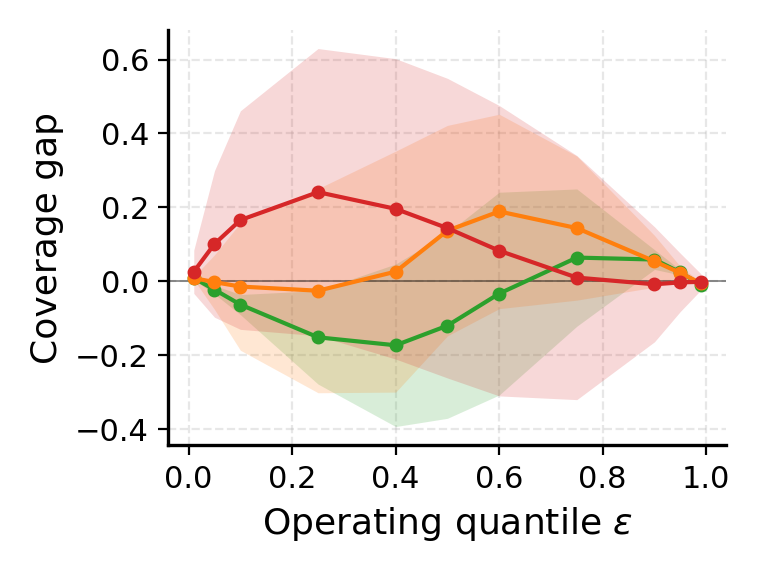}
        \caption{Coverage gap vs.\ group.}
        \label{fig:km_tailgap}
    \end{subfigure}
    \caption{Predictive accuracy (a,b) and calibration (c,d) by volatility group (\grouplegend).}
    \label{fig:km_acc}
\end{figure*}

We evaluate the \ac{KM} along three complementary dimensions. First, accuracy measures how closely the predicted capacity matches the realized capacity. Second, uncertainty calibration to evaluate whether predictive intervals achieve the true capacity as often as they claim. Finally, the sensitivity to the number of posterior samples used during inference. 

We compare the proposed \ac{BNN} with three forecasting baselines: a deterministic \ac{LSTM} (i.e., producing a single value with no uncertainty), Monte Carlo (MC) Dropout (i.e., approximating Bayesian inference by sampling dropout masks)~\cite{gal2016dropout}, and a deep ensemble (i.e., reading uncertainty from the spread of independently trained networks)~\cite{lakshminarayanan2017}. All models predict the one-second-ahead schedulable capacity $C_k(t+h)$ on the held-out test day.

\subsubsection{Accuracy}

Fig.~\ref{fig:km_acc}(a)-(b) report point accuracy. The deep ensemble achieves the lowest median \ac{MAE} ($20$~TOPS), followed by the deterministic \ac{LSTM} ($21$~TOPS), whereas MC Dropout and the proposed \ac{BNN} obtain comparable errors ($35$ and $36$~TOPS, respectively). Point accuracy alone, therefore, does not distinguish the \ac{BNN}; its advantage lies in predictive calibration, which is discussed next. Forecasting errors are concentrated in the \textit{Volatile} BS group, where median \ac{MAE} increases from tens to hundreds of TOPS, confirming that the proposed grouping captures forecasting difficulty. The \ac{LSTM} is particularly sensitive to abrupt changes, with the worst case (BS C672) reaching $19\,725$~TOPS because it cannot express predictive uncertainty.

Figure~\ref{fig:km_mape} reports the corresponding relative error. Although \textit{Volatile} BSs exhibit much larger absolute errors, they also provide substantially larger compute capacity, keeping the \ac{BNN}'s mean absolute percentage error nearly constant across all groups ($1.1\%$, $1.2\%$, and $1.0\%$ for \textit{Stable}, \textit{Moderate}, and \textit{Volatile}, respectively). Since admission decisions depend on relative rather than absolute capacity error, this explains why the \textit{Volatile} group ultimately achieves admission performance closest to the oracle (\S\ref{sec:eval_dm}).
\textit{As a conclusion, point accuracy alone does not distinguish the probabilistic models; however, the consistently low relative error across regimes ensures that forecasting errors have limited impact on admission decisions, which depend on capacity ratios rather than absolute values.} 

\subsubsection{Uncertainty Calibration}
While accuracy measures how close the predicted capacity is to the true capacity, the calibration asks a different question: \textit{do predictive intervals contain the real capacity as often as they claim?} 

\noindent\textbf{Reliability.} Figure~\ref{fig:km_reliability} reports empirical versus nominal (real) coverage, where perfect calibration lies on the diagonal. The proposed \ac{BNN} closely follows this perfect case across the full range, achieving $95.7\%$ empirical coverage at the nominal 95\% level (IQR $95.3$--$96.3\%$). In contrast, MC Dropout ($99.0\%$) and the deep ensemble ($98.1\%$) consistently over-cover, indicating overly conservative predictive intervals.

This behavior is reflected in the calibration error at the $95\%$ level, defined as the absolute difference between empirical and nominal coverage. The \ac{BNN} achieves the lowest median error ($0.79$ percentage points), compared with $3.11$ for the ensemble and $4.03$ for MC Dropout. \textit{The \ac{BNN} provides the best balance between predictive sharpness and calibration.}



\noindent\textbf{Tail calibration.} 
Nominal $95\%$ coverage alone is insufficient because \ac{SMART} bases admission decisions on lower predictive quantiles \eqref{eq:saa_cap}. Fig.~\ref{fig:km_tailgap}  reports the coverage gap (empirical minus nominal) across the full quantile range. A gap near zero indicates that the realized admission risk matches the target risk. \textit{Stable} and \textit{Volatile} \acp{BS} exhibit opposite behaviors. \textit{Stable} \acp{BS} are over-confident, with the largest negative gap ($-0.17$ at $\varepsilon=0.40$), whereas \textit{Volatile} \acp{BS} are under-confident, with the largest positive gap ($+0.24$ at $\varepsilon=0.25$).
\textit{These results indicates that a single calibration temperature cannot calibrate all volatility regimes simultaneously.}


\begin{table}[!htbp]
\centering
\caption{\ac{ECE} between empirical and nominal coverage, with sharpness
(MPIW$_{95}$ in TOPS).}
\label{tab:calibration}
\footnotesize
\setlength{\tabcolsep}{4pt}
\begin{tabular}{lccccc}
\toprule
 & \multicolumn{4}{c}{\ac{ECE} (pp)} & Sharpness \\
\cmidrule(lr){2-5}
Model & Stable & Moderate & Volatile & All & MPIW$_{95}$ \\
\midrule
BNN  & $11.8$ & $14.1$ & $12.4$ & $13.2$ & $287$ \\
MC Dropout  & $19.6$ & $19.0$ & $21.2$ & $19.8$ & $1006$ \\
Ensemble  & $19.0$ & $17.9$ & $13.3$ & $18.0$ & $313$ \\
\bottomrule
\end{tabular}
\end{table}

\noindent\textbf{Overall calibration.} 
Table~\ref{tab:calibration} presents the \ac{ECE}, calculated as the mean absolute difference between empirical and nominal coverage for the 50\%, 80\%, 90\%, and 95\% prediction intervals, along with the median width of the 95\% prediction interval (MPIW$_{95}$).
The \ac{BNN} achieves the lowest \ac{ECE} overall ($13.2$~pp, versus $18.0$ for the ensemble and $19.8$ for MC Dropout) while maintaining the narrowest prediction intervals ($287$~TOPS, versus $313$ and $1006$~TOPS). 
It consistently outperforms the competing probabilistic models across all volatility groups. 
Residual calibration error is concentrated at intermediate coverage levels because temperature scaling is optimized for the $95\%$ interval. 
Nevertheless, the \ac{BNN}'s advantage is statistically significant across the $45$ BSs: a Friedman test rejects equal calibration error among the probabilistic models ($\chi^2=42.2$, $p<10^{-9}$), while Holm-corrected Wilcoxon signed-rank tests confirm lower calibration error than both MC Dropout ($p<2\times10^{-4}$) and the deep ensemble ($p<5\times10^{-3}$). \textit{The \ac{BNN} provides the most reliable predictive quantiles while avoiding the excessive conservatism exhibited by alternative probabilistic predictors.}

\subsubsection{Sample-count sensitivity}

Finally, we evaluate the sensitivity of the \ac{KM} to the number of Monte Carlo samples. We vary the sample size $S$ in \eqref{eq:bnn_predictive} over $S\in\{25,\dots,1000\}$ across all $45$ BSs. Point accuracy is insensitive to $S$, while calibration improves with larger sample sizes. 
In particular, on the \textit{Volatile} group, the spread of the $95\%$ predictive coverage (\ac{PICP}$_{95}$) across BSs and MC trials decreases from $14.3$~pp at ($S=25$) to $1.9$~pp at ($S=1000$).
However, admission decisions stabilize at much smaller sample sizes. 
At $S=100$, re-running the \ac{SAA} and \ac{CVaR} sweeps changes the median deviation from the oracle by at most $0.02$ percentage points compared to $S=1000$. \textit{As a conclusion, stable admission decisions can be obtained with only $S=100$ posterior samples, providing a favorable trade-off between decision robustness and inference latency.}

 \vspace{0.2cm}
\greybox{\textbf{Findings:} \textit{The \ac{BNN} combines competitive point accuracy with substantially better calibration than alternative probabilistic predictors, while requiring only a modest number of posterior samples for stable inference. These properties make it well suited to drive \ac{SMART}'s uncertainty-aware admission policy, whose performance depends on reliable predictive quantiles rather than point forecasts alone. }}
 

\subsection{DM: Admission Performance}
\label{sec:eval_dm}

Unless otherwise stated, all reported quantities are expressed as deviations from the oracle admission. A value of zero, therefore, indicates that the admission policy matches the oracle, which has perfect knowledge of the future capacity. The risk parameters $\varepsilon$ (\ac{SAA}) and $\alpha$ (\ac{CVaR}) control the conservativeness of the admission decision: increasing either parameter relaxes the admission constraint and allows more tasks to be admitted.

\subsubsection{Risk--Admission Tradeoff} 
\begin{figure}[t]
    \centering
    \begin{subfigure}[t]{0.49\columnwidth}
        \centering
        \includegraphics[width=\linewidth]{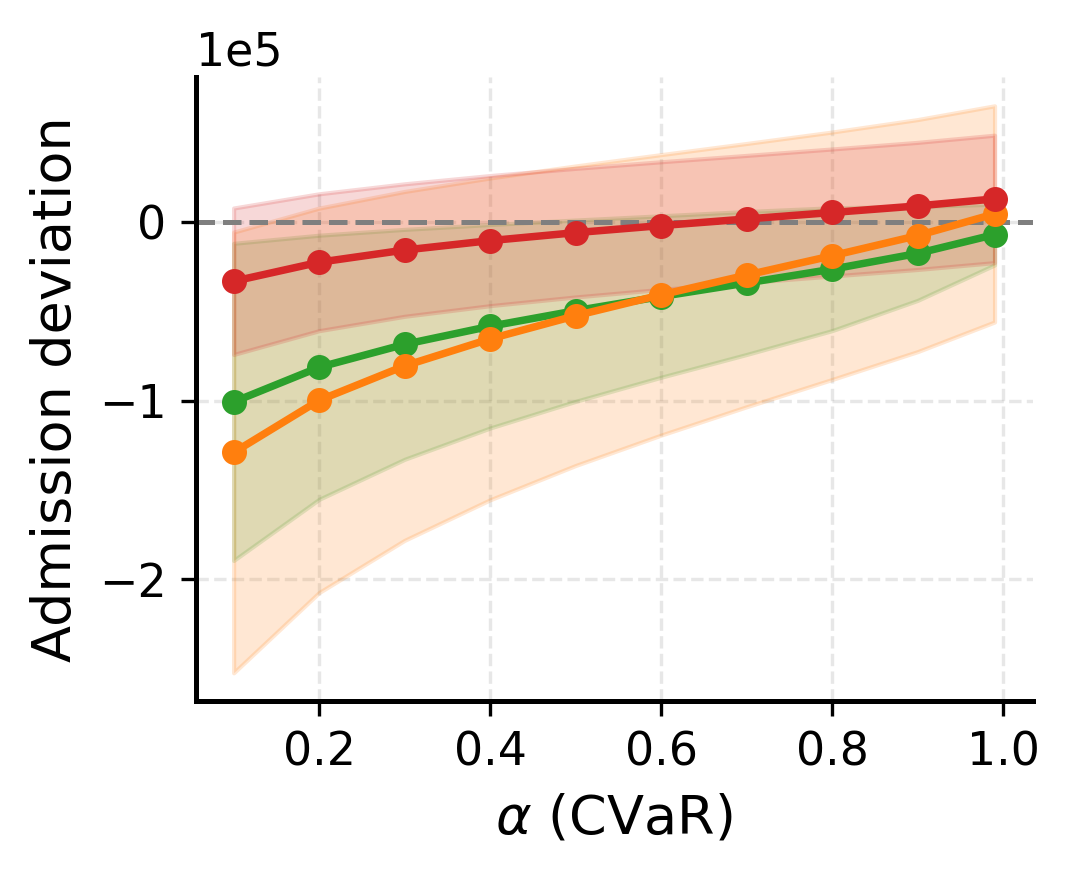}
        \caption{Admission vs.\ $\alpha$ (CVaR).}
        \label{fig:dm_adm_alpha}
    \end{subfigure}
    \hfill
    \begin{subfigure}[t]{0.49\columnwidth}
        \centering
        \includegraphics[width=\linewidth]{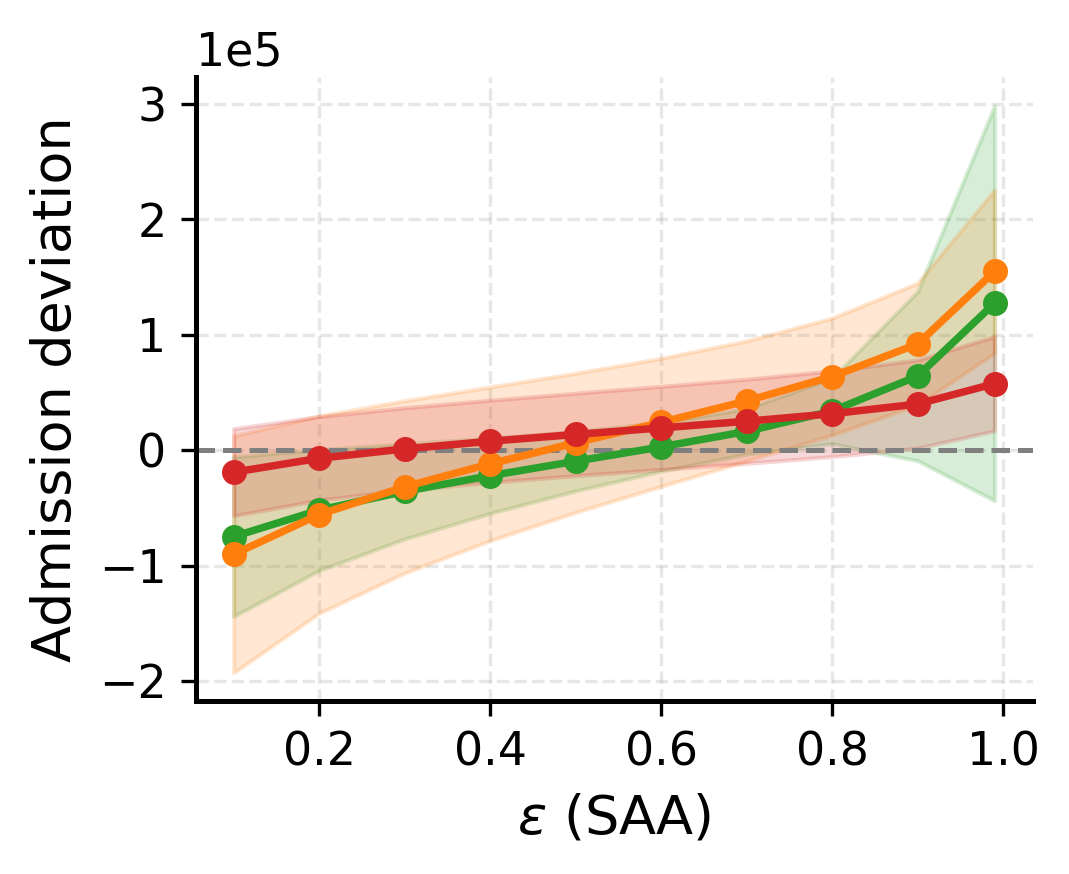}
        \caption{Admission vs.\ $\varepsilon$ (SAA).}
        \label{fig:dm_adm_eps}
    \end{subfigure}
    \caption{Admission deviation against the oracle, by group. \grouplegend.}
    \label{fig:dm_groups}
\end{figure}

Figure~\ref{fig:dm_groups} reports the admission deviation only with respect to the oracle for the three volatility groups. As we can observe, the behavior of \ac{SAA} is more aggressive. Small values of $\varepsilon$ lead to conservative admission, while larger values eventually cause over-admission. At $\varepsilon=0.99$, the \textit{Stable} and \textit{Moderate} groups exceed the oracle admission, indicating that optimistic capacity estimates can accept more load than the realized capacity can support. On the other hand, the \textit{Volatile} group stays nearest the oracle across the whole $\varepsilon$ range, since, against its large capacity, the same prediction error affects a smaller fraction of its admission.

Under \ac{CVaR}, all groups under-admit at small values of $\alpha$ and gradually approach the oracle as the risk level increases. The \textit{Volatile} group remains closest to the oracle over the entire range and slightly exceeds it at large $\alpha$ values. \textit{Stable} and \textit{Moderate} cells approach the oracle more conservatively. 

Finally, we observe that, for both approaches, the optimal risk parameter configuration ($\varepsilon^\star$ for SAA and $\alpha^\star$ for CVaR) depends on the volatility regime of each cell. We can see that the \textit{Volatile} group favors lower risk parameters and a more conservative approach. On the other hand,  \textit{Stable} and \textit{Moderate} cells require larger risk parameters to recover the oracle admission. This observation suggests that the admission risk needs to be adjusted to each cell's volatility. \textit{As a conclusion, both \ac{SAA} and \ac{CVaR} achieve admission levels close to the oracle while offering different risk profiles. \ac{SAA} provides more aggressive admission, whereas \ac{CVaR} offers a safer and more conservative behavior. The optimal risk level depends on the uncertainty regime of the cell.}

\subsubsection{Impact of Uncertainty Decomposition}

\begin{figure}[t]
    \centering
    \begin{subfigure}[t]{0.32\columnwidth}
        \centering
        \includegraphics[width=\linewidth]{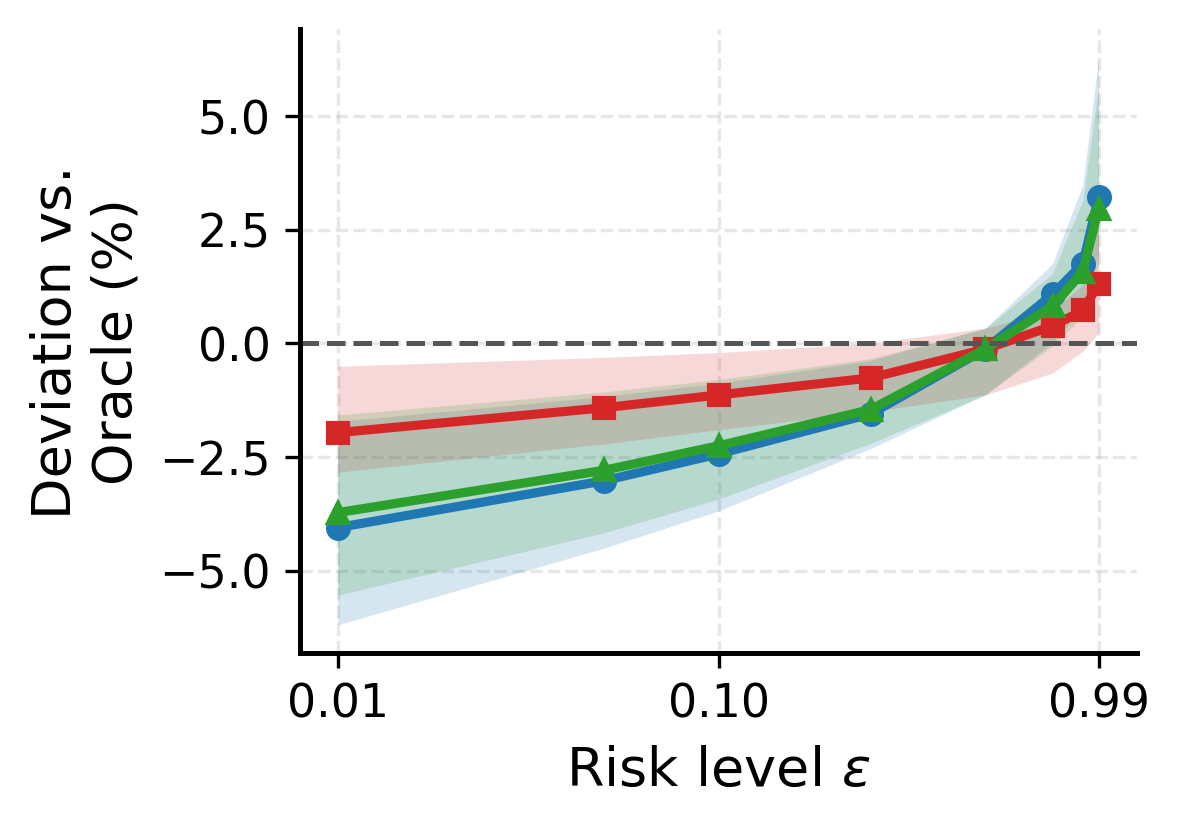}
        \caption{Stable.}
        \label{fig:abl_stable}
    \end{subfigure}
    \hfill
    \begin{subfigure}[t]{0.32\columnwidth}
        \centering
        \includegraphics[width=\linewidth]{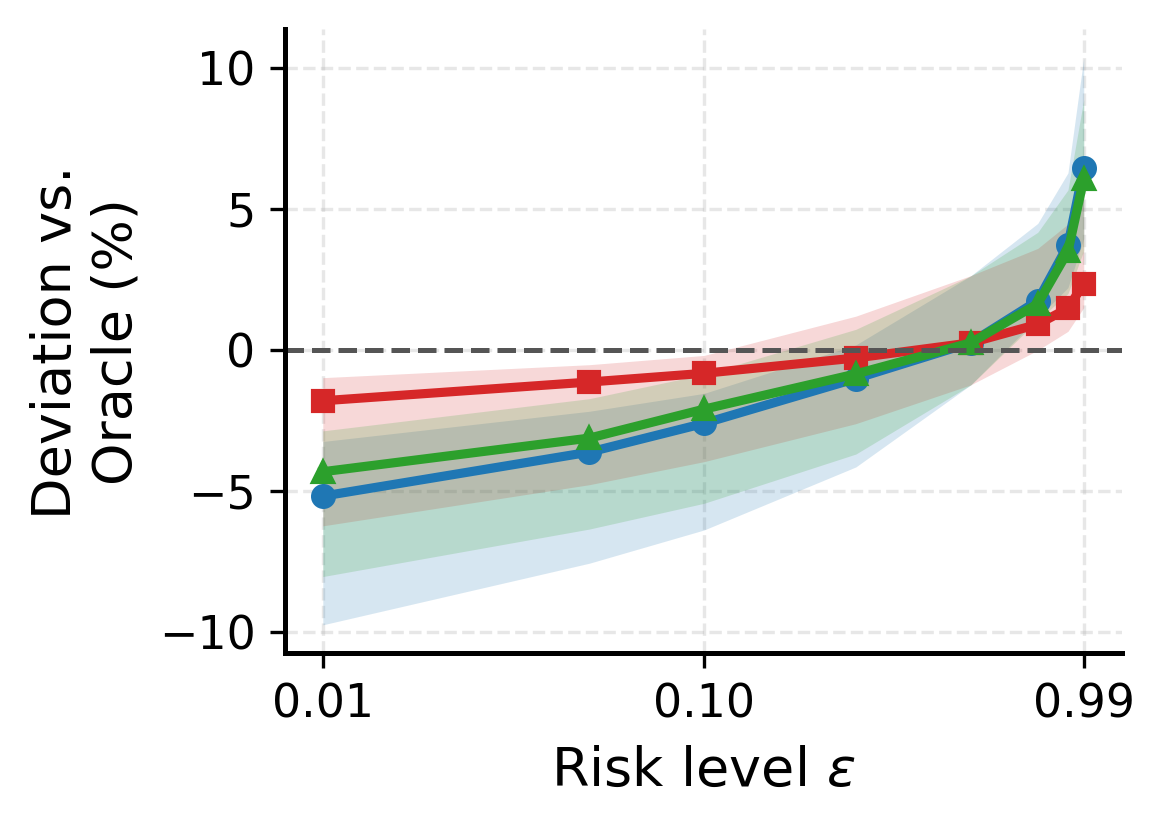}
        \caption{Moderate.}
        \label{fig:abl_moderate}
    \end{subfigure}
    \hfill
    \begin{subfigure}[t]{0.32\columnwidth}
        \centering
        \includegraphics[width=\linewidth]{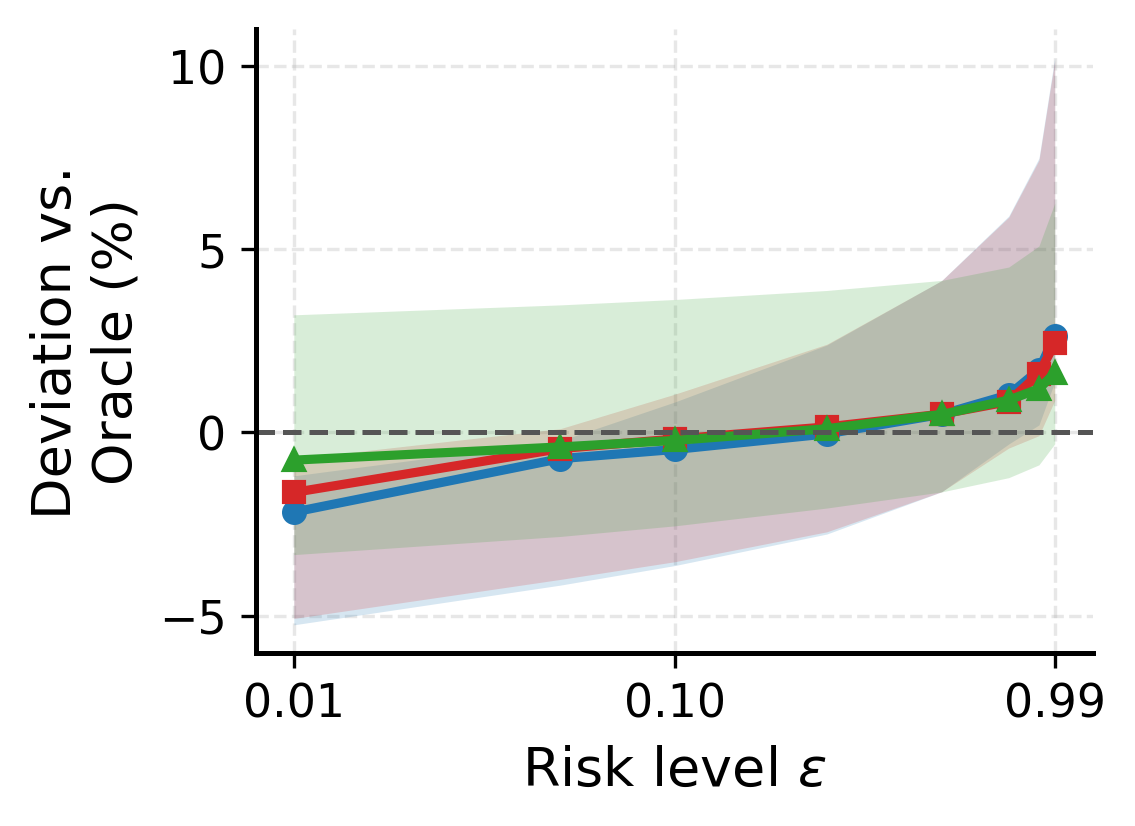}
        \caption{Volatile.}
        \label{fig:abl_volatile}
    \end{subfigure}
    \caption{Variance decomposition ablation.\\ \ablationlegend.}
    \label{fig:abl}
\end{figure}

Figure~\ref{fig:abl} evaluates the contribution of epistemic and aleatoric uncertainty to the admission decision. We considered three variants. The first variant, noted \textbf{TOTAL}, uses the full predictive draws. The seconde variant, noted \textbf{EPI\textsubscript{only}}, keeps the posterior-mean samples $\mu^{(s)}(x_t)$ alone. Finally, the last variant, noted \textbf{ALE\textsubscript{only}}, uses the posterior mean plus fresh draws from $\mathcal{N}(0,\hat\sigma_{\mathrm{ale}})$. 
The results show that the two uncertainty components do not contribute equally to the admission decision. In \textit{Stable} and \textit{Moderate} cells, the aleatoric component dominates the lower predictive tail and largely determines the effective capacity budget. In contrast, the epistemic component becomes increasingly important in the \textit{Volatile} group. \textit{In conclusion, removing either component changes admission behavior, and the direction of that change depends on the regime rather than simply on shrinking or expanding a uniform safety margin}.
\vspace{.2cm}
\greybox{\textit{
\textbf{Findings.} Epistemic and aleatoric uncertainty influence admission decisions differently across mobility regimes. While aleatoric uncertainty primarily determines the effective admission budget, the contribution of epistemic uncertainty increases in highly volatile cells, making their explicit separation beneficial for risk-aware admission.
}}

\subsection{Comparison Baselines}
\label{sec:eval_baselines}

Beyond the oracle comparison, we evaluate \ac{SMART} against seven baselines spanning three classes, all evaluated on the same $45$ BSs and workload (Table~\ref{tab:baselines}). These baselines allow us to isolate the benefits of forecasting (Class~A vs.~Class~B), of distributional information (Class~B vs.~Class~C), and of the proposed probabilistic forecasters (B1 vs.~B2):

\begin{itemize}[leftmargin=*]
    \item[-] \textbf{Class~A~-} Reactive policies that do not forecast capacity: A1 admits a task if its demand fits within the last observed capacity (adapted from~\cite{feraudo2023}), while A2a and A2b greedily prioritize tasks by decreasing (First-Fit Decreasing) and increasing demand, respectively.
    \item[-] \textbf{Class~B~-} Policies using point forecasts of mean future capacity: B1 relies on the predictive mean of the proposed \ac{BNN}, whereas B2 uses a deterministic \ac{LSTM}.
    \item[-] \textbf{Class~C~-} Uncertainty-aware policies: C1 applies split-conformal prediction using the per-BS residual quantile \(q_\varepsilon\), while C2 adopts robust optimization with a box uncertainty set (\(\mu-\Gamma\sigma\)).
\end{itemize}
Table~\ref{tab:baselines} compares \textit{Admission} and \textit{Capacity-Violation} rates across the three BS volatility groups and baselines. A \textit{Capacity-Violation} occurs when the admitted load exceeds the real capacity during a one-second decision window.

\begin{table*}[t]
\centering
\caption{Mean $\pm$ $95\%$ confidence interval of admission and capacity-violation rates across the $45$ cells.}
\label{tab:baselines}
\footnotesize
\setlength{\tabcolsep}{2.5pt}
\resizebox{\linewidth}{!}{%
\begin{tabular}{l cccc cccc}
\toprule
& \multicolumn{4}{c}{Admission (\%)} & \multicolumn{4}{c}{Cap.\ violation (\%)} \\
\cmidrule(lr){2-5} \cmidrule(lr){6-9}
Policy & Stable & Moderate & Volatile & All & Stable & Moderate & Volatile & All \\
\midrule
A1\,~Reactive (Feraudo) & 75.5\,$\pm$\,2.3 & 75.7\,$\pm$\,1.3 & 78.6\,$\pm$\,0.2 & 76.6\,$\pm$\,0.9 & 0.1\,$\pm$\,0.1 & 0.0\,$\pm$\,0.1 & 0.0\,$\pm$\,0.0 & 0.0\,$\pm$\,0.0 \\
A2a~Greedy FFD & 3.7\,$\pm$\,0.0 & 3.7\,$\pm$\,0.1 & 3.7\,$\pm$\,0.0 & 3.7\,$\pm$\,0.0 & 0.0\,$\pm$\,0.1 & 0.0\,$\pm$\,0.0 & 0.0\,$\pm$\,0.0 & 0.0\,$\pm$\,0.0 \\
A2b~Greedy SFD & 26.1\,$\pm$\,0.3 & 26.1\,$\pm$\,0.2 & 26.5\,$\pm$\,0.0 & 26.3\,$\pm$\,0.1 & 0.0\,$\pm$\,0.0 & 0.0\,$\pm$\,0.0 & 0.0\,$\pm$\,0.0 & 0.0\,$\pm$\,0.0 \\
B1\,~Mean only (BNN) & 75.4\,$\pm$\,2.4 & 75.9\,$\pm$\,1.7 & 79.1\,$\pm$\,0.8 & 76.8\,$\pm$\,1.0 & 0.2\,$\pm$\,0.3 & 0.1\,$\pm$\,0.1 & 0.0\,$\pm$\,0.0 & 0.1\,$\pm$\,0.1 \\
B2\,~Mean only (LSTM) & 75.6\,$\pm$\,2.2 & 75.7\,$\pm$\,1.3 & 74.1\,$\pm$\,3.6 & 75.1\,$\pm$\,1.4 & 0.9\,$\pm$\,1.8 & 0.4\,$\pm$\,0.6 & 0.0\,$\pm$\,0.0 & 0.4\,$\pm$\,0.6 \\
C1\,~Conformal  & 75.4\,$\pm$\,2.4 & 75.9\,$\pm$\,1.7 & 79.0\,$\pm$\,0.8 & 76.8\,$\pm$\,1.0 & 0.2\,$\pm$\,0.2 & 0.1\,$\pm$\,0.1 & 0.0\,$\pm$\,0.0 & 0.1\,$\pm$\,0.1 \\
C2\,~Robust & 74.9\,$\pm$\,2.5 & 75.2\,$\pm$\,1.8 & 78.7\,$\pm$\,0.7 & 76.3\,$\pm$\,1.1 & 0.1\,$\pm$\,0.1 & 0.1\,$\pm$\,0.1 & 0.0\,$\pm$\,0.0 & 0.1\,$\pm$\,0.0 \\
\midrule
SMART-SAA ($\varepsilon^\star$) & 93.5\,$\pm$\,3.4 & 93.8\,$\pm$\,1.7 & 97.9\,$\pm$\,0.4 & 95.1\,$\pm$\,1.3 & 4.9\,$\pm$\,1.1 & 5.1\,$\pm$\,2.5 & 8.2\,$\pm$\,4.7 & 6.1\,$\pm$\,1.7 \\
SMART-CVaR ($\alpha^\star$) & 93.4\,$\pm$\,3.4 & 93.7\,$\pm$\,1.9 & 97.7\,$\pm$\,0.4 & 95.0\,$\pm$\,1.3 & 5.1\,$\pm$\,3.0 & 4.9\,$\pm$\,1.5 & 2.5\,$\pm$\,1.2 & 4.2\,$\pm$\,1.1 \\
oracle (capacity) & 93.5\,$\pm$\,3.4 & 94.0\,$\pm$\,1.7 & 97.9\,$\pm$\,0.4 & 95.1\,$\pm$\,1.3 & \textemdash & \textemdash & \textemdash & \textemdash \\
\bottomrule
\end{tabular}}
\end{table*}

For \ac{SMART}, each BS operates at its calibrated risk level $\varepsilon^\star$ for \ac{SAA} or $\alpha^\star$ for \ac{CVaR}), selected such that its admission rate does not exceed the oracle. At these operating points, \ac{SMART}-\ac{SAA} and \ac{SMART}-\ac{CVaR} admit $95.1\%$ and $95.0\%$ of tasks, respectively (cf. \textit{All} column), effectively matching the oracle ($95.1\%$). In contrast, the remaining baselines admit only $75\%$ to $77\%$ of tasks, except the greedy policies A2a ($3.7\%$) and A2b ($26.3\%$). 

Across all policies, \textit{Volatile} BSs achieve the highest admission rates, including under the oracle ($97.9\%$ versus $93.5\%$ for \textit{Stable} BSs), indicating that admission performance is primarily driven by available capacity dynamics rather than forecasting difficulty. 
Neither reactive policies nor methods that rely solely on point forecasts or generic uncertainty estimates match \ac{SMART}'s performance, indicating that the gain comes from translating predictive uncertainty into a risk-aware admission budget.


The \ac{SMART}'s higher admission rate comes at the cost of more frequent \textit{Capacity Violations}. Overall (cf. \textit{All} column), \ac{SMART}-\ac{SAA} and \ac{SMART}-\ac{CVaR} violate capacity in $6.1\%$ and $4.2\%$ of admitted decisions, respectively, whereas all baselines remain below 0.5\% by admitting substantially fewer tasks. \ac{CVaR} consistently provides the safer operating point, particularly for \textit{Volatile} BSs (i.e., $2.5\%$ of violated decisions versus $8.2\%$ for \ac{SAA}), because averaging the lower $\alpha$-tail produces a more conservative capacity estimate than the single \ac{SAA} quantile. 

%% file: 06_conclusions.tex
\section{Conclusions and Future Work}
\label{sec:conclusion}
We presented \ac{SMART}, an \ac{ETSI}-compatible mechanism for uncertainty-aware admission of opportunistic vehicular compute. \ac{SMART} is composed of two modules: a Knowledge Manager that learns a calibrated predictive distribution of future per-BS compute capacity using a \ac{BNN}, and a Decision Manager that performs risk-aware admission via chance-constrained optimization using \ac{SAA} and \ac{CVaR} approximations. 

Using a real-world taxi mobility dataset, we showed that the proposed \ac{BNN} provides the best-calibrated forecasts at the nominal $95\%$ coverage level. Under the compute-only admission model, \ac{SMART} admits approximately $95\%$ of tasks (i.e., $95.1\%$ with \ac{SAA} and $95.0\%$ with \ac{CVaR}), matching the Oracle and outperforming seven reactive, point-forecast, and uncertainty-aware baselines by approximately $23\%$ in relative admission rate 

This improvement is achieved at capacity-violation rates of $6.1\%$ (\ac{SAA}) and $4.2\%$ (\ac{CVaR}). After accounting for violations, the fraction of tasks successfully served remains between $89\%$ and $91\%$, compared with about $77\%$ for the best baseline. All decisions are executed within the one-second control interval with lightweight per-BS control-plane overhead. 

We analyze the operating conditions under which \ac{SMART} remains effective. Sensitivity to variability in base-station compute availability is significant. Consequently, under abrupt capacity drops, aggressive admission reduces available headroom, suggesting the need for complementary mechanisms such as reserved capacity, proactive migration, or fallback to static \ac{MEC} resources. Overall, the  homogeneous-capacity assumption used in this work isolates mobility-driven uncertainty and should be interpreted as an evaluation setting rather than a deployment guarantee. Future work will focus on online risk adaptation, multi-step forecasting, and extending the model to include communication and queueing effects beyond compute-only resources.